\documentclass[sigconf]{acmart}

\usepackage{booktabs} % For formal tables

\setcopyright{rightsretained}
\usepackage[english]{babel}
\usepackage{subcaption}
\usepackage[inline]{enumitem}
\usepackage{amsmath}
\usepackage{stmaryrd}
\usepackage{graphicx}
\usepackage{subcaption}
\usepackage{multirow}

\newenvironment{ppl}{\fontfamily{ppl}\selectfont}{}

\def\ignore#1{}

\begin{document}
\title{Response Ranking with Deep Matching Networks and External Knowledge in Information-seeking Conversation Systems}
%\subtitle{[Model Representation and Inference Note]
%\titlenote{This technical report will be further revised as a SIGIR 2018 full paper submission.}

% Option A: Multi-Turn Response Ranking with Deep Matching Networks and External Knowledge in Retrieval-based Conversation System
% Option B: Response Ranking with Deep Matching Networks and External Knowledge in Information-seeking Conversation System (Focus on multi-turn QA conversations)
% Option C: Response Ranking with Deep Matching Networks and External Knowledge in Retrieval-Based Conversation System
% Option D: Response Ranking with Deep Matching Networks and External Knowledge in QA focused Conversation System

%\author{Anonymous Authors}
\author{Liu Yang$^1$ \quad  Minghui Qiu$^2$ \quad Chen Qu $^1$ \quad  Jiafeng Guo$^3$ \quad Yongfeng Zhang$^4$ \quad W. Bruce Croft$^1$ \quad Jun Huang$^2$ \quad Haiqing Chen$^2$}

\affiliation{%
	\institution{
	$^1$ Center for Intelligent Information Retrieval, University of Massachusetts Amherst \quad
	$^2$ Alibaba Group  \\ % \quad
	$^3$ Institute of Computing Technology, Chinese Academy of Sciences \quad
	$^4$ Dept. of Computer Science, Rutgers University}
	%\streetaddress{P.O. Box 1212}
}
\email{{lyang, chenqu, croft}@cs.umass.edu, {minghui.qmh, huangjun.hj, haiqing.chenhq}@alibaba-inc.com}
\email{guojiafeng@ict.ac.cn, yongfeng.zhang@rutgers.edu}
%	\institution{

%	\institution{
%	$^1$ University of Massachusetts Amherst \quad
%	$^2$ Alibaba Group \quad
%	$^3$ Chinese Academy of Sciences \quad
%	$^4$ Rutgers University}

%	\institution{
%	$^1$ Center for Intelligent Information Retrieval, University of Massachusetts Amherst, MA, USA \\
%	$^2$ Alibaba Group, Zhejiang, China \\
%	$^3$  Institute of Computing Technology,
%	Chinese Academy of Sciences, Beijing, China \\
%	$^4$ Department of Computer Science, Rutgers University, NJ, USA}

%\author{}
% CAS Key Lab of Network Data Science and Technology,

\begin{abstract}
	\noindent Intelligent personal assistant systems with either text-based or voice-based conversational interfaces are becoming increasingly popular around the world. Retrieval-based conversation models have the advantages of returning fluent and informative responses. Most existing studies in this area are on open domain ``chit-chat'' conversations or task / transaction oriented conversations. More research is needed for information-seeking conversations. There is also a lack of modeling external knowledge beyond the dialog utterances among current conversational models. In this paper, we propose a learning framework on the top of deep neural matching networks that leverages external knowledge for response ranking in information-seeking conversation systems. We  incorporate external knowledge into deep neural models with pseudo-relevance feedback and QA correspondence knowledge distillation. Extensive experiments with three information-seeking conversation data sets including both open benchmarks and commercial data show that, our methods outperform various baseline methods including several deep text matching models and the state-of-the-art method on response selection in multi-turn conversations. We also perform analysis over different response types, model variations and ranking examples. Our models and research findings provide new insights on how to utilize external knowledge with deep neural models for response selection and have implications for the design of the next generation of information-seeking conversation systems.
\end{abstract}

% We no longer use \terms command

% \keywords{\noindent Conversational Models; Deep Learning; Question Answering}

%\copyrightyear{2018}
%\acmYear{2018} 
%\setcopyright{acmcopyright}
%\acmConference{SIGIR'18}{}{July 8-12, 2018, Ann Arbor, Michigan, U.S.A.  }\acmPrice{15.00}\acmDOI{http://dx.doi.org/xx}
%\acmISBN{xxxx}
%%\acmConference{SIGIR’17}{}{August 7--11, 2017, Shinjuku, Tokyo, Japan}

\copyrightyear{2018} 
\acmYear{2018} 
\setcopyright{acmcopyright}
\acmConference[SIGIR '18]{The 41st International ACM SIGIR Conference on Research and Development in Information Retrieval}{July 8--12, 2018}{Ann Arbor, MI, USA}
\acmBooktitle{SIGIR '18: The 41st International ACM SIGIR Conference on Research and Development in Information Retrieval, July 8--12, 2018, Ann Arbor, MI, USA}
\acmPrice{15.00}
\acmDOI{10.1145/3209978.3210011}
\acmISBN{978-1-4503-5657-2/18/07}

\fancyhead{}
\settopmatter{printacmref=true, printfolios=false}

\maketitle

\section{Introduction}
\label{sec:intro}

% Present in a logical way

%1. What is the task and Why it is important
%The recent boom of artificial intelligence has witnessed the

%The fast development of artificial intelligence contributes to the emerging and flourishing of many intelligent personal assistant systems, including Amazon Alexa, Apple Siri, Alibaba AliMe, Microsoft Cortana and Google Now, among wider population around the world. 

Personal assistant systems, such as Apple Siri, Google Now, Amazon Alexa, and Microsoft Cortana, are becoming ever more widely used\footnote{For example, over 100M installations of Google Now (Google, \url{http://bit.ly/1wTckVs}); 15M sales of Amazon Echo (GeekWire, \url{http://bit.ly/2xfZAgX}); more than 141M monthly users of Microsoft Cortana (Windowscentral, \url{http://bit.ly/2Dv6TVT}).}. These systems, with either text-based or voice-based conversational interfaces, are capable of voice interaction, information search, question answering and voice control of smart devices. This trend has led to an interest in developing conversational search systems, where users would be able to ask questions to seek information with conversation interactions. Research on speech and text-based conversational search has also recently attracted significant attention in the information retrieval (IR) community.

%Both speech and text based conversational search systems are emerging topics among the information retrieval (IR) community in recent years\footnote{\url{https://sites.google.com/view/cair-ws/}}\footnote{ \url{http://scai.info/}}, attracting enormous attention of researchers from both academia and industry. 

%2. What is the problems/missing parts of current state-of-art work

% Review current main research on generation based and retreival based methods for conversation system. Retreivel based methods have the advantages of ...
% For retreivel based methods, most current methods are on single-turn conversation, e.g. Wang et al. 2013

Existing approaches to building conversational systems include generation-based methods \cite{DBLP:conf/emnlp/RitterCD11,DBLP:conf/acl/ShangLL15} and retrieval-based methods \cite{DBLP:journals/corr/JiLL14,DBLP:conf/sigir/YanSW16,DBLP:conf/cikm/YanSZW16,DBLP:conf/sigir/YanZE17}. Compared with generation-based methods, retrieval-based methods have the advantages of returning fluent and informative responses. Most work on retrieval-based conversational systems studies response ranking for \textit{single-turn conversation} \cite{DBLP:conf/emnlp/WangLLC13}, which only considers a current utterance for selecting responses. Recently, several researchers have been studying \textit{multi-turn conversation} \cite{DBLP:conf/sigir/YanSW16,DBLP:conf/emnlp/ZhouDWZYTLY16,DBLP:conf/acl/WuWXZL17,DBLP:conf/sigir/YanZE17}, which considers the previous utterances of the current message as the conversation context to select responses by jointly modeling context information, current input utterance and response candidates. However, existing studies are still suffering from the following weaknesses: 
%  Due to the rapid growth of user generated conversation data on social platforms ( e.g., Twitter, Facebook, Weibo) and community question answering sites (e.g. Quora, Stack Overflow, Yahoo Answers), it is become increasingly practical as the increased amount of data makes it more likely that a suitable response will be found.  
%  although achieved great progress on building human computer conversation systems, 
% \yfcomment{We may be claiming too much for retrieval-based approaches, this may annoy a reviewer if he happens to work on generation-based approaches}.

% Most recently, some researchers studied multi-turn conversation. e.g, Yan et al. DL2R, Tian et al. Zhou et al. Multi-View, Wu et al. SMN.  However, the current research and methods in this area have a few weakness:
% 2.1  Most current works on multi-turn conversation are mainly looking at open domain chit-chat conversations
% 2.2 Most currrent open domain conversation models are merely model the matching patterns between the conversational context with response candidates ignoring external knowledge beyond the dialog. Whether external knowledge is useful for conversatioin resonponse ranking and how to do this effectively have not been well studied.

%\begin{enumerate}
(1) \textbf{Most existing studies are on open domain chit-chat conversations or task / transaction oriented conversations}. Most current work \cite{DBLP:conf/emnlp/RitterCD11,DBLP:conf/acl/ShangLL15, DBLP:journals/corr/JiLL14,DBLP:conf/sigir/YanSW16,DBLP:conf/cikm/YanSZW16,DBLP:conf/sigir/YanZE17} is looking at open domain chit-chat conversations as in microblog data like Twitter and Weibo. There is some research on task oriented conversations \cite{Young:2010:HIS:1621140.1621240,wen2016network,bordes2017learning}, where there is a clear goal to be achieved through conversations between the human and the agent. However, the typical applications and data are related to completing transactions like ordering a restaurant or booking a flight ticket. Much less attention has been paid to \textit{information oriented conversations}, which is referred to as \textit{information-seeking conversations} in this paper. Information-seeking conversations, where the agent is trying to satisfy the information needs of the user through conversation interactions, are closely related to conversational search systems. More research is needed on response selection in information-seeking conversation systems. 
%\ylcomment{will try to give a clearer definition of ``information-seeking conversations''}%\yfcomment{Are the concepts `transaction oriented conversations', `information oriented conversations', and `information-seeking conversations' already existing concepts or newly created concepts in this paper? If already existing, better give reference, otherwise, we should give a clearer definition. Besides, `information oriented conversations' and `information-seeking conversations' are repetitions of the same concept, usually a principle is to define as few terminologies as possible and be consistent throughout the text.}

(2) \textbf{Lack of modeling external knowledge beyond the dialog utterances}. Most research on response selection in conversation systems are purely modeling the matching patterns between user input message (either with context or not) and response candidates, which ignores external knowledge beyond the dialog utterances. Similar to Web search, information-seeking conversations could be associated with massive external data collections that contain rich knowledge that could be useful for response selection. This is especially critical for information-seeking conversations, since there may be not enough signals in the current dialog context and candidate responses to discriminate a good response from a bad one due to the wide range of topics for user information needs. An obvious research question is how to utilize external knowledge effectively for response ranking. This question has not been well studied, despite the potential benefits for the development of information-seeking conversation systems. 

%\yfcomment{For (1) and (2), I think the discussion of chit-chat systems can be moved to related work; after introducing multi-turn conversation in the second paragraph, we can only briefly introduce chit-chat and use it to introduce the concept of information-seeking conversation, and then quickly go to the key of the paper -- using external knowledge for info-seeking conversation.}

%\end{enumerate}

%3.What is your contribution/findings/ the novel parts of the model or archtecture/  summary of experiment results

To address these research issues, we propose a learning framework on top of deep neural matching networks that leverages external knowledge for response ranking in information-seeking conversation systems. We study two different methods on integrating external knowledge into deep neural matching networks as follows: %To this end, we created a new data set MSDialog, which contains more than $35,000$ technical support dialogs on Microsoft products from the Microsoft answers Community\footnote{\url{answers.microsoft.com}}. We perform fine-grained user intent annotation and analysis with MSDialog and another popular benchmark data Ubuntu Dialog Corpus (UDC) \cite{DBLP:journals/corr/LowePSP15}, which consists of one million two-person technical support dialogs. The data analysis results show that most utterances from these two data sets are on questions, potential answers, information requests and user feedback, which represent typical content in information-seeking conversations. With these two conversation data, we build learning models for response ranking.
%Unlike many previous studies on microblog conversation data, we are looking at effective methods for response selection in information-seeking conversations.

% \yfcomment{We do not have to talk about so much details of data preparation in introduction, just briefly introduce the data with a few sentences, and leave the details to Section 3.}
%  (\ylcomment{add percentage here later}) 
% 3.1 Focuse on information-seeking conversations, which is different from previous works on chit-chat Twitter data
% In this paper, we are looking at information-seeking conversations, which is targeting on satisifying user information needs with conversation interactions. 
% To this end, we create a data set MSDialog...

% 3.2 Incorporate external knowledge in neural models to enchance response ranking performance
% This is especially critical for information-seeking converstaions since ... To show this, we give an example dialog in Table ...

% There are serveral chanllenges on incorporating external knowledge into response ranking in conversational system. e.g. (naive solutions won't work) it is very challenge to directly extract responses from external collections. For example,...  We did the following to overcome these chanllenges

%\begin{itemize} \end{itemize}
(1) \textbf{Incorporating external knowledge via pseudo-relevance feedback}. Pseudo-relevance feedback (PRF) has been proven effective in improving the performance of many retrieval models \cite{Lavrenko:2001:RBL:383952.383972,Lv:2009:CSM:1645953.1646259,Zamani:2016:PFB:2983323.2983844,Zhai:2001:MFL:502585.502654,rocchio71relevance,Cao:2008:SGE:1390334.1390377}. 
The motivation of PRF is to assume a certain number of top-ranked documents from the initial retrieval run to be relevant and use these feedback documents to improve the original query representation. For conversation response ranking, many candidate responses are much shorter compared with conversation context,  which could have negative impacts on deep neural matching models. Inspired by the key idea of PRF, we propose using the candidate response as a query to run a retrieval round on a large external collection. Then we extract useful information from the (pseudo) relevant feedback documents to enrich the original candidate response representation. 

%Important terms will be extracted to expand the candidate response and then the deep matching networks match the conversation context with expanded candidate responses. 
%  In Web search, users often issue very short queries, which lacks important terms to better represent user information seeking intent, leading to poor retrieval results. PRF is an effective technique to get better query representations. 

(2) \textbf{Incorporating external knowledge via QA correspondence knowledge distillation}. Previous neural ranking models enhanced the performance of retrieval models such as BM25 and QL, which mainly rely on lexical match information, via modeling semantic match patterns in text \cite{Guo:2016:DRM:2983323.2983769,DBLP:conf/cikm/HuangHGDAH13,Mitra:2017:LMU:3038912.3052579}. For response ranking in information-seeking conversations, the match patterns between candidate responses and conversation context can be quite different from the well studied lexical and semantic matching. Consider the following sample utterance and response from the conversations in the Microsoft Answers community \footnote{\url{https://answers.microsoft.com/}} shown in Table \ref{tab:qapost_example}. A Windows user proposed a question about the windows update failure on ``restart install''. An expert replied with a response pointing to a potential cause ``Norton leftovers''. The match signals between the problem ``restart install'' and the cause ``Norton leftovers'' may not be captured by simple lexical and semantic matching. To derive such match patterns, we need to rely on external knowledge to distill QA correspondence information. We propose to extract the ``correspondence'' regularities between question and answer terms from retrieved external QA pairs. We define this type of match patterns as a ``\textit{correspondence match}'', which will be incorporated into deep matching networks as external knowledge to help response selection in information-seeking conversations. 

%\yfcomment{Also, because the introduction is already very long, we'd better shorten (1) and (2) here, only introduce the key idea, and leave the details to the main text.}

% https://answers.microsoft.com/en-us/windows/forum/windows_10-update-winpc/windows-update-failure/b3ac8dbf-4ad1-4431-b31d-e67185f91851

\begin{table}
	\scriptsize
	%\fontfamily{ppl}\selectfont
	\begin{ppl}
		\caption{Sample utterance and response from the conversations in the Microsoft Answers community. This figure could be more readable with color print. Note that the purpose of this figure is to illustrate examples and differences among these three types of matches instead of exhaustively labeling all three types of matches between the two texts.}
		\vspace{-0.1in}
		 %\ylcomment{Will update this table later to highlight the QA term co-occurrences in the retrieved QA pairs from external collection.  Then link this match information back to the words in utterances/responses in the dialog
		\begin{tabular}{|p{8.0cm}|}
			\hline
			\multicolumn{1}{|l|}{\emph{QA Dialog Title: }: \textbf{Windows Update Failure }}\\
			\multicolumn{1}{|l|}{\emph{Dialog Tags}: \textbf{Windows, Windows 10, Windows update, recovery, backup, PC}}\\
		    \underline{USER:} I have \textcolor{blue}{Windows10}, version 1511, OS Build 10586.1106. For the past year I have tried to \textcolor{magenta}{upgrade} from this without success. \textcolor{magenta}{Upgrade} download OK but on installing only get to 85 - 93\% and then on \textcolor{red}{restart install} previous version of \textcolor{blue}{windows} (the 1511 version), I have  \textcolor{blue}{Windows} update assistant installed. Any help or advice on this would be most welcome. \\
			David\\
			\hline
			\multicolumn{1}{|l|}{\emph{Responses}} %& \multicolumn{1}{c|}{Votes~(\textcolor{blue})}\\
			\\
			\hline
			\underline{AGENT: James (Microsoft MVP - Windows Client) }:\newline
			\textbf{Response}:There's not a doubt in my mind that those \textcolor{red}{Norton ``leftovers''} is your troublemaker here - but now that the \textcolor{red}{Norton Removal Tool} has been deprecated and especially since the new-fangled \textcolor{red}{Norton Remove and Reinstall tool} doesn't get rid of the \textcolor{red}{leftovers}, a manual \textcolor{magenta}{upgrade} or a clean install of \textcolor{blue}{Microsoft Win10} appears to be your only possible resolution here. Feel free to give \textcolor{red}{Norton/Symantec} a piece of your mind! \\
			\hline
			Term Match: \textcolor{magenta}{Magenta} \qquad  Semantic Match:  \textcolor{blue}{Blue} \qquad    Correspondence Match:  \textcolor{red}{Red}\\
			\hline
		\end{tabular} \label{tab:qapost_example}
	\end{ppl}
	\vspace{-0.3cm}
\end{table}

We conduct extensive experiments with three information-seeking conversation data sets: the MSDialog data which contains crawled customer service dialogs from Microsoft Answers community , a popular benchmark data Ubuntu Dialog Corpus (UDC) \cite{DBLP:journals/corr/LowePSP15}, and another commercial customer service data AliMe from Alibaba group. We compare our methods with various deep text matching models and the state-of-the-art baseline on response selection in multi-turn conversations. Our methods outperform all baseline methods regrading a variety of metrics. % for response ranking in information-seeking conversations.

To sum up, our contributions can be summarized as follows:
% The novel properties of the proposed model and our contribution can be summerized as follows:

%\begin{enumerate}
 (1) \textbf{Focusing on information-seeking conversations and building a new benchmark data set.} We target information-seeking conversations to push the boundaries of conversational search models. To this end, we create a new information-seeking conversation data set MSDialog on technical support dialogs of Microsoft products and released it to the research community \footnote{The MSDialog dataset can be downloaded from ~\url{https://ciir.cs.umass.edu/downloads/msdialog}.  We also released our source code at ~\url{https://github.com/yangliuy/NeuralResponseRanking} .}.

 (2) \textbf{Integrating external knowledge into deep neural matching networks for response ranking.}  We propose a new response ranking paradigm for multi-turn conversations by incorporating external knowledge into the matching process of dialog context and candidate responses. Under this paradigm, we design two different methods with pseudo relevance feedback and QA correspondence knowledge distillation to integrate external knowledge into deep neural matching networks for response ranking.
	
(3) \textbf{Extensive experimental evaluation on benchmark / commercial data sets and promising results.} Experimental results with three different information-seeking conversation data sets show that our methods outperform various baseline methods including the state-of-the-art method on response selection in multi-turn conversations. We also perform analysis over different response types, model variations and ranking examples to provide insights.
%\end{enumerate}

%Our contributions can be summarized as follows:
%%\vspace{-1.5cm}
%% Highlight contributions/ what we have done and the differences between our work with previous works

%4.Roadmap: the structure of the whole paper
%\textbf{Roadmap}. 
%The rest of our paper is organized as follows. We will review related work in ...

\section{Related Work}
\label{sec:rel}

% The google doc link for the related work survey
% Refer to the realted work section of the NEUIR17 paper and SIE paper

%\begin{table*}[th]
%	\centering
%	\begin{footnotesize}
%		\caption{Summary of experimental datasets in this paper.}
%		\vspace{-0.15in}
%		\label{tab:data_summery}
%		\begin{tabular}{ l|| l l l}
%			\hline \hline
%			Data & QA Dialog (QAD) & External QA Collection (QAC) & Source  \\
%			\hline 
%			UDC & Ubuntu QA dialogs extracted from IRC network & AskUbuntu data dump & Both QAD and QAC are publicly available
%			\\
%			MSDialog & Customer service dialogs on Microsoft products & Stack Overflow data dump & QAD is crawled from Web; QAC is publicly available \\
%			E-commerce & Customer service dialogs from a large eCommerce company & Internal product FAQ data & Both QAD and QAC are proprietary data \\
%			\hline \hline
%		\end{tabular}
%	\end{footnotesize}
%\end{table*} 

Our work is related to research on conversational search, neural conversational models and neural ranking models.

%\textbf{Information-seeking Conversations}

\textbf{Conversational Search.}
Conversational search has received significant attention with the emerging of conversational devices in the recent years. Radlinski and Craswell described the basic features of conversational search systems \cite{radlinski2017theoretical}. Thomas et al. \cite{thomas2017misc} released the Microsoft Information-Seeking Conversation (MISC) data set, which contains information-seeking conversations with a human intermediary, in a setup designed to mimic software agents such as Siri or Cortana. But this data is quite small (in terms of the number of dialogs) for the training of neural models. Based on state-of-the-art advances on machine reading, Kenter and de Rijke \cite{kenter-attentive-2017} adopted a conversational search approach to question answering. Except for conversational search models, researchers have also studied the medium of conversational search. Arguello et al. \cite{arguello2017factors} studied how the medium (e.g., voice interaction) affect user requests in conversational search. Spina et al. studied the ways of presenting search results over speech-only channels to support conversational search \cite{spina2017extracting,trippas2015towards}. Yang et al. \cite{DBLP:journals/corr/YangZZGC17} investigated predicting the new question that the user will ask given the past conversational context. Our research targets at the response ranking of information-seeking conversations, with deep matching networks and integration of external knowledge.

% Vakulenko et al. \cite{vakulenko2017conversational} adopted interactive storytelling as a tool to enable exploratory search within a conversational interface.

\textbf{Neural Conversational Models.}
Recent years there are growing interests on research about conversation response generation and ranking with deep learning and reinforcement learning \cite{DBLP:conf/acl/ShangLL15,DBLP:conf/sigir/YanSW16,DBLP:conf/cikm/YanSZW16,DBLP:conf/sigir/YanZE17,DBLP:conf/acl/LiGBSGD16,DBLP:conf/emnlp/LiMRJGG16,DBLP:conf/naacl/SordoniGABJMNGD15,bordes2017learning}. Existing work includes retrieval-based methods \cite{DBLP:conf/acl/WuWXZL17,DBLP:conf/emnlp/ZhouDWZYTLY16,DBLP:conf/sigir/YanSW16,DBLP:conf/sigir/YanZE17,DBLP:journals/corr/JiLL14,DBLP:journals/corr/LowePSP15} and generation-based methods \cite{DBLP:conf/acl/ShangLL15,DBLP:conf/acl/TianYMSFZ17,DBLP:conf/emnlp/RitterCD11,DBLP:conf/naacl/SordoniGABJMNGD15,DBLP:journals/corr/VinyalsL15,DBLP:conf/emnlp/LiMRJGG16,bordes2017learning,P17-1045,alime-chat}. Sordoni et al. \cite{DBLP:conf/naacl/SordoniGABJMNGD15} proposed a neural network architecture for response generation that is both context-sensitive and data-driven utilizing the Recurrent Neural Network Language Model architecture. %Bordes et al. \cite{bordes2017learning} proposed a testbed to break down the strengths and shortcomings of end-to-end dialog systems in goal-oriented applications based on Memory Networks  \cite{weston2015memory,sukhbaatar2015end,graves2014neural}. Shang et al. \cite{DBLP:conf/acl/ShangLL15} proposed Neural Responding Machine (NRM), which is a RNN encoder-decoder framework for short text conversation and showed that it outperformed retrieved-based methods and SMT-based methods for single round conversation.
 Our work is a retrieval-based method. There are some research on multi-turn conversations with retrieval-based method. Wu et al. \cite{DBLP:conf/acl/WuWXZL17} proposed a sequential matching network that matches a response with each utterance in the context on multiple levels of granularity to distill important matching information. The main difference between our work with their research is that we consider external knowledge beyond dialog context for multi-turn response selection. We show that incorporating external knowledge with pseudo-relevance feedback and QA correspondence knowledge distillation is important and effective for response selection.

\textbf{Neural Ranking Models.}
Recently a number of neural ranking models have been proposed for information retrieval, question answering and conversation response ranking. These models could be classified into three categories \cite{Guo:2016:DRM:2983323.2983769}. The first category is the representation focused models. These models will firstly learn the representations of queries and documents separately and then calculate the similarity score of the learned representations with functions such as cosine, dot, bilinear or tensor layers. A typical example is the DSSM \cite{DBLP:conf/cikm/HuangHGDAH13} model, which is a feed forward neural network with a word hashing phase as the first layer to predict the click probability given a query string and a document title. The second category is the interaction focused models, which build a query-document term pairwise interaction matrix to capture the exact matching and semantic matching information between the query-document pairs. Then the interaction matrix will be fed into deep neural networks which could be CNN \cite{DBLP:conf/nips/HuLLC14,DBLP:conf/aaai/PangLGXWC16,alime-tl}, term gating network with histogram or value shared weighting mechanism \cite{Guo:2016:DRM:2983323.2983769,Yang:2016:ARS:2983323.2983818} to generate the final ranking score. In the end, the neural ranking models in the third category combine the ideas of the representation focused models and interaction focused models to joint learn the lexical matching and semantic matching between queries and documents \cite{Mitra:2017:LMU:3038912.3052579,alime-tl}. The deep matching networks used in our research belong to the interaction focused models due to their better performances on a variety of text matching tasks compared with representation focused models \cite{DBLP:conf/nips/HuLLC14,DBLP:conf/aaai/PangLGXWC16,Guo:2016:DRM:2983323.2983769,Yang:2016:ARS:2983323.2983818,DBLP:conf/acl/WuWXZL17, Xiong:2017:ENA:3077136.3080809}. We study different ways to build the interaction matching matrices to capture the matching patterns in term spaces, sequence structures and external knowledge signals between dialog context utterances and response candidates.

\section{Our Approach}
\subsection{Problem Formulation}
The research problem of response ranking in information-seeking conversations is defined as follows. We are given an information-seeking conversation data set $\mathcal{D} = \{(\mathcal{U}_i, \mathcal{R}_i,  \mathcal{Y}_i)\}_{i=1}^N$, where $ \mathcal{U}_i = \{u_i^1, u_i^2, \dots, u_i^{t-1} , u_i^t\} $ in which $ \{u_i^1, u_i^2, \dots, u_i^{t-1}\} $ is the dialog context and $u_i^t$ is the  input utterance in the $t$-th turn. $\mathcal{R}_i$ and $\mathcal{Y}_i$ are a set of response candidates $ \{r_i^1, r_i^2, \dots, r_{i}^k\}_{k=1}^M $ and the corresponding binary labels $ \{y_i^1, y_i^2, \dots, y_{i}^k\} $, where $y_i^k=1$ denotes $r_i^k$ is a true response for $\mathcal{U}_i$. Otherwise $y_i^k=0$. In order to integrate external knowledge, we are also given an external collection  $\mathcal{E}$, which is related to the topics discussed in conversation $\mathcal{U}$. Our task is to learn a ranking model $f(\cdot)$ with $\mathcal{D}$ and $\mathcal{E}$. For any given $\mathcal{U}_i$, the model should be able to generate a ranking list for the candidate responses $\mathcal{R}_i$ with $f(\cdot)$.  The external collection $\mathcal{E}$ could be any massive text corpus. In our paper, $\mathcal{E}$ are historical QA posts in Stack Overflow data dump \footnote{\url{https://stackoverflow.com/}} for MSDialog, AskUbuntu data dump  \footnote{\url{https://askubuntu.com/}} for Ubuntu Dialog Corpus and product QA pairs for AliMe data. % from Alibaba group. %The summary of experimental data sets used in this paper is shown in Table \ref{tab:data_summery}.

  \begin{table}[!t]
  	\footnotesize
	\caption{A summary of key notations in this work. Note that all vectors are denoted with bold cases.}
	\vspace{-10pt}
	\begin{tabular}
		{ p{0.06\textwidth} | p{0.38\textwidth}} \hline  \hline
		$\mathcal{D}$ & The conversation data set used for training/validation/testing\\\hline
		$\mathcal{E}$ & The collection for the retrieval and distillation of external knowledge \\ \hline
		$u_i^t, \mathcal{U}_i, \mathcal{U}$ & The $t$-th utterance of the $i$-th dialog, all utterances of the $i$-th dialog and the set of all dialog utterances\\\hline
		$r_i^k, \mathcal{R}_i, \mathcal{R}$ & The $k$-th response candidate for the  $i$-th dialog, all response candidates of the  $i$-th dialog and the set of all candidate responses\\\hline
		$r_i^{k'}$ & The $k$-th expanded response candidate for the  $i$-th dialog \\ \hline
		$y_i^k, \mathcal{Y}$ & The label for the $k$-th response candidate for the $i$-th dialog and the set of all labels \\ \hline
		$f(\cdot)$ & The ranking model learnt with $\mathcal{D}$ and $\mathcal{E}$ \\ \hline
		$f( \mathcal{U}_i, r_i^k ) $ & The predicted matching score between $\mathcal{U}_i$ and $r_i^k$ \\ \hline
		$N$ & The total number of dialogs in $\mathcal{D}$\\ \hline
		$M$ & The total number of response candidates for $\mathcal{U}_i$\\ \hline
		$W$ & The number of expanded words in response candidates\\ \hline
		$\theta$ & The language model constructed from the pseudo relevance feedback document set for response candidate expansion\\ \hline
		$P, \mathcal{P} $ & The number of top ranked QA posts  retrieved from $\mathcal{E}$ and  the top ranked QA post set \\ \hline
		$l_r, l_u$ & The length of a response candidate and the length of an utterance \\ \hline
		$d$ & The number of dimensions of word embedding vectors \\ \hline
		$\mathbf{M}_1$,$\mathbf{M}_2$,$\mathbf{M}_3$ & Interaction matrices between dialog utterance $u_i^t$ and candidate response $r_i^k$ or  $r_i^{k'}$ for word embedding similarity, sequence hidden representation similarity and QA  correspondence matching similarity\\ \hline
		$m_{1,i,j}$ & The $(i,j)$-th element in the interaction matrix $\mathbf{M}_1$\\ \hline
		$c$ & The window size for the utterances in dialog context, which is the maximal number of previous utterances modeled \\ \hline  \hline
	\end{tabular}\label{tab:notation}
	\vspace{-10pt}
\end{table}

% Present in a mathmatical way
\subsection{Method Overview}
 In the following sections, we describe the proposed learning framework built on the top of deep matching networks and external knowledge for response ranking in information-seeking conversations. A summary of key notations in this work is presented in Table \ref{tab:notation}. In general, there are three modules in our learning framework:
 
 %\begin{enumerate}
 (1) \textbf{Information retrieval (IR) module:} Given the information seeking conversation data $\mathcal{D}$ and external QA text collection $\mathcal{E}$, this module is to retrieve a small relevant set of QA pairs $\mathcal{P}$ from $\mathcal{E}$ with the response candidate $\mathcal{R}$ as the queries. These retrieved QA pairs $\mathcal{P}$ become the source of external knowledge. % to be integrated into the neural models.
 	
 (2) \textbf{External knowledge extraction (KE) module:} Given the retrieved QA pairs $\mathcal{P}$  from the IR module, this module will extract useful information as term distributions, term co-occurrence matrices or other forms as external knowledge.
 	
 (3) \textbf{Deep matching network (DMN) module:} This is the module to model the extracted external knowledge from $\mathcal{P}$ , dialog utterances $ \mathcal{U}_i$ and the response candidate $ r_i^k$ to learn the matching pattern, over which it will accumulate and predict a matching score $f( \mathcal{U}_i, r_i^k ) $ for $ \mathcal{U}_i$ and $ r_i^k$. %The response candidates $\mathcal{R}$  will be ranked according to these matching scores.
% \end{enumerate}
 
\begin{figure*}[th]
 	\center
 	\includegraphics*[viewport=0mm 0mm 280mm 90mm, scale=0.50]{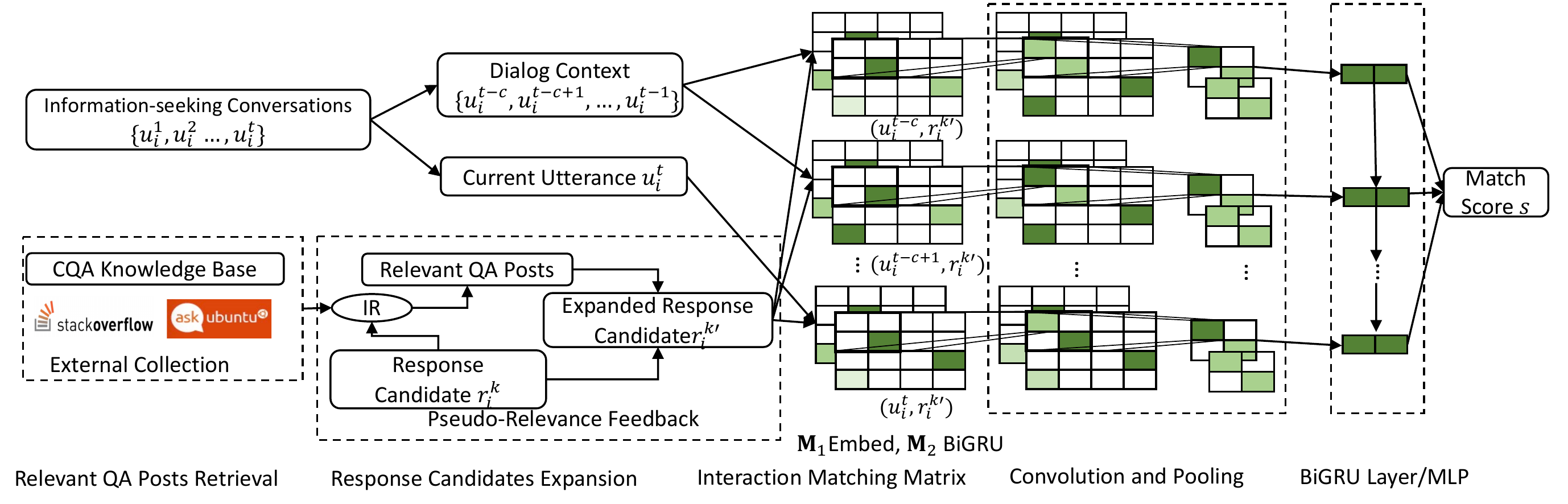}
 	\vspace{-0.4cm}
 	\caption{The architecture of DMN-PRF model for conversation response ranking.}\label{fig:dmn-prf}
 	%\vspace{-0.4cm}
 \end{figure*}

 We explore two different implementations under this learning framework as follows: 1) Incorporating external knowledge into deep matching networks via pseudo-relevance feedback (DMN-PRF). The architecture of DMN-PRF model is presented in Figure \ref{fig:dmn-prf}. 2) Incorporating external knowledge via QA correspondence knowledge distillation (DMN-KD). The architecture of DMN-KD model is presented in Figure \ref{fig:dmn-kd}. We will present the details of these two models in Section \ref{sec:method_dmn_prf} and Section \ref{sec:method_dmn_kd}.

 %\vspace{-0.3cm}

%\ylcomment{Until 19:47 am 01/29/2018, 5th round of checking/updating to here.}

\subsection{Deep Matching Networks with Pseudo-Relevance Feedback}
\label{sec:method_dmn_prf}
% Present in a mathmatical way

%\ylcomment{Until 19:25 01/24/2018, double check and revise the paper/reduce the length to here.}
\subsubsection{\textbf{Relevant QA Posts Retrieval}} %\linebreak
We adopt different QA text collections for different conversation data (e.g. Stack Overflow data for MSDialog, AskUbuntu for UDC). The statistics of these external collections are shown in Table \ref{tab:ext_collection_statistics}. We download the data dumps for Stack Overflow and AskUbuntu from archive.org\footnote{\url{https://archive.org/download/stackexchange}}. We index the QA posts in Stack Overflow in most recent two years and all the QA posts in AskUbuntu. Then we use the response candidate $ r_i^k$ as the query to retrieve top $P$ \footnote{In our experiments, we set $P=10$.} QA posts with BM25 as the source for external knowledge. %  Then we use these indexes to perform QA posts retrieval in the next step for response candidate expansion.

\begin{table}
	\centering
	\footnotesize
	\caption{Statistics of external collections for QA pairs retrieval and knowledge extraction. Note that ``\#QWithAcceptedA'' means  ``number of questions with an accepted answer''. The other names use similar abbreviations.}
	\vspace{-0.1in}
	\label{tab:ext_collection_statistics}
	\begin{tabular}{l|l|l}
		\hline \hline
		Collection Name     & SOTwoYears & AskUbuntu \\ \hline  \hline
		StartDate           & 12/4/2015    & 7/28/2010   \\ \hline
		EndDate             & 9/1/2017     & 9/1/2017    \\ \hline
		\#QAPosts           & 9,563,530  & 629,198   \\ \hline
		\#Time              & 2 Years    & 7 years   \\ \hline
		XMLFileDiskSize     & 17GB       & 799MB     \\ \hline
		\#Question          & 4,188,937  & 271,233   \\ \hline
		%\#QWithTitle & 4,188,937  & 271,233   \\ \hline
		%\#QWithAcceptedA    & 1,751,787  & 92,259    \\ \hline
		%\#QWithAtLeastOneA  & 3,178,814  & 213,830   \\ \hline
		\%QWithAcceptedA    & 41.82\%    & 34.01\%   \\ \hline
		\%QWithAtLeastOneA  & 75.89\%    & 78.84\%   \\ \hline   \hline
		%\#Answer            & 5,362,237  & 357,965   \\ \hline  \hline
		%\#AWithTitle   & 296        & 0         \\ \hline \hline
	\end{tabular}
\end{table}
%\vspace{-0.1in}

%\ylcomment{Until 14:53pm, 3rd round of checking and revising to here.}
\subsubsection{\textbf{Candidate Response Expansion}} %\linebreak
\label{sec:response_expansion}
The motivation of Pseudo-Relevance Feedback (PRF) is to extract terms from the top-ranked documents in the first retrieval results to help discriminate relevant documents from irrelevant ones \cite{Cao:2008:SGE:1390334.1390377}. The expansion terms are extracted either according to the term distributions (e.g. extract the most frequent terms) or extracted from the most specific terms (e.g. extract terms with the maximal IDF weights) in feedback documents. Given the retrieved top QA posts $\mathcal{P}$ from the previous step, we compute a language model $\theta = P(w| \mathcal{P})$ using $\mathcal{P}$. Then we extract the most frequent $W$\footnote{In our experiments, we set $W = 10$. }  terms from $\theta$ as expansion terms for response candidate $r_i^k$ and append them at the end of $r_i^k$. For the query $r_i^k$, we perform several preprocessing steps including tokenization, punctuation removal and stop words removal.  QA posts in both Stack Overflow and AskUbuntu have two fields: ``Body'' and ``Title''. We choose to search the ``Body'' field since we found it more effective in experiments. 

% Only report the best setting of DMN-PRF, which is DMN-PRF-Body
%  \footnote{According to the statistics in Table \ref{tab:ext_collection_statistics}， all the question posts have ``Body'' and ``Title'' fields. Only a few answer posts in Stack Overflow have the ``Title'' field. }
% all the question posts have ``Body'' and ``Title'' fields. Only a few answer posts in Stack Overflow have the ``Title'' field. }, we generate two versions of retrieval results for candidate response expansion, which is corresponding to searching the ``Body'' or ``Title'' field of the QA posts. We choose to re
%We refer the models using these two different retrieval strategies as ``DMN-PRF-Body'' and ``DMN-PRF-Title'' respectively. We will compare the effectiveness of these two retrieval strategies in Section \ref{sec:exps}.

\subsubsection{\textbf{Interaction Matching Matrix}}
% Present in a mathmatical way
The expanded response candidates and dialog contexts will be modeled by a deep neural matching network. Given an expanded response $r_i^{k'}$ and an utterance $u_i^t$ in the context $\mathcal{U}_{i}$, the model firstly looks up a global embedding dictionary to represent  $r_i^{k'}$ and $u_i^t$ as two sequences of embedding vectors $\mathbf{E}(r_i^{k'})=[\mathbf{e}_{r,1}, \mathbf{e}_{r,2}, \cdots, \mathbf{e}_{r,l_r}]$ and  $\mathbf{E}(u_i^t)=[\mathbf{e}_{u,1}, \mathbf{e}_{u,2}, \cdots, \mathbf{e}_{u,l_u}]$, where $\mathbf{e}_{r,i} \in \mathbb{R}^d$, $\mathbf{e}_{u,i} \in \mathbb{R}^d$  are the embedding vectors of the $i$-th word in $r_i^{k'}$ and $u_i^t$ respectively. Given these two word embedding sequences, there are two different methods to learn matching patterns: representation focused methods and interaction focused methods \cite{Guo:2016:DRM:2983323.2983769}. Here we adopt the interaction focused methods due to their better performances over a number of text matching tasks \cite{DBLP:conf/nips/HuLLC14,DBLP:conf/aaai/PangLGXWC16,DBLP:conf/aaai/WanLGXPC16,Yang:2016:ARS:2983323.2983818}. Specifically, the model builds two interaction matrices with $\mathbf{E}(r_i^{k'}) \in \mathbb{R}^{d\times l_r}$ and $\mathbf{E}(u_i^t) \in \mathbb{R}^{d\times l_u}$: a word pairwise similarity matrix $\mathbf{M}_1$ and a sequence hidden representation similarity matrix $\mathbf{M}_2$.  $\mathbf{M}_1$ and $\mathbf{M}_2$ will be two input channels of a convolutional neural network (CNN) to learn important matching features, which will be aggregated by the final BiGRU layer and a multi-layer perceptron (MLP) to generate a matching score.
% \ylcomment{In the current implementation, these two matrices are the dot-product matrices between word embedding and sequence representations learned in BiGRU. I will try more variations like indicator, cosine, bi-linear or single direction GRU, etc later.}

Specifically, in the input channel one, $\forall i, j$, the element $m_{1,i,j}$ in the $\mathbf{M}_1$ is defined by $m_{1,i,j} = \mathbf{e}_{r,i}^T \cdot \mathbf{e}_{u,j}$.
%\begin{footnotesize}	
%\begin{equation} \label{Eqn:matrix1_element_ij}
%\end{equation}
%\end{footnotesize}
$\mathbf{M_1}$ models the word pairwise similarity between $r_i^{k'}$ and $u_i^t$ via the dot product similarity between the embedding representations.

For input channel two, we firstly employ bidirectional gated recurrent units (BiGRU) \cite{DBLP:journals/corr/ChungGCB14} to encode $r_i^{k'}$ and $u_i^t$ into two hidden representations. A BiGRU consists two GRUs that run in opposite directions on sequence $\mathbf{E}(r_i^{k'})$: a forward GRUs processing the sequence as it is ordered, and another backward GRUs processing the sequence in its reverse order. These two GRUs will generate two sequences of hidden states $(\vec{\mathbf{h}_1}, \cdots, \vec{\mathbf{h}_{l_r}})$ and $(\overset{{}_{\shortleftarrow}}{\mathbf{h}_1}, \cdots, \overset{{}_{\shortleftarrow}}{\mathbf{h}_{l_r}}  )$. BiGRU then concatenates the forward and the backward hidden states to form the final hidden vectors for $r_i^{k'}$ as $\mathbf{h}_i = {[\vec{\mathbf{h}_i}, \overset{{}_{\shortleftarrow}}{\mathbf{h}_i} ]}_{i=1}^{l_r}$. More specifically, $\forall i$, the hidden state vector $\vec{\mathbf{h}_i}\in \mathbb{R}^O$ is calculated by the following formulas:

% Similar to the LSTM unit \cite{Hochreiter:1997:LSM:1246443.1246450}, the GRU has gating units that modulate the flow of information inside the unit, however, without having a separate memory cells. It has similar performances with the LSTM unit on sequence modeling, but with lower computational costs. 
 \begin{footnotesize}
\begin{equation}\label{Eqn:gru}
\begin{aligned}
\mathbf{z}_i&=\sigma(\mathbf{W}_z\mathbf{e}_{r,i}+\mathbf{U}_z\vec{\mathbf{h}}_{i-1}+\mathbf{b}_z)\\
\mathbf{r}_i&=\sigma(\mathbf{W}_r\mathbf{e}_{r,i}+\mathbf{U}_r \vec{\mathbf{h}}_{i-1}+\mathbf{b}_r)\\
\tilde{\mathbf{h}}_i&=\tanh(\mathbf{W}_h\mathbf{e}_{r,i}+\mathbf{U}_h(\mathbf{r}_i\circ\vec{\mathbf{h}}_{i-1})+\mathbf{b}_h)\\
\vec{\mathbf{h}}_i&=(\mathbf{1}-\mathbf{z}_i)\circ\vec{\mathbf{h}}_{i-1}+\mathbf{z}_i\circ\tilde{\mathbf{h}}_i
\end{aligned}
\end{equation}
\end{footnotesize}
where $\mathbf{z}_i$ and $\mathbf{r}_i$ are an update gate and a reset gate respectively. $\textbf{e}_{r,i}, \vec{\textbf{h}}_i$ are the input and hidden state output of the network at time step $i$. $\mathbf{W}_z,\mathbf{W}_r,\mathbf{W}_h,\mathbf{U}_z,\mathbf{U}_r,\mathbf{U}_h$ and $\mathbf{b}_z,\mathbf{b}_r,\mathbf{b}_h $ are parameter matrices and bias vectors to be learned. The backward hidden state $ \overset{{}_{\shortleftarrow}}{\mathbf{h}_i} \in \mathbb{R}^O$ is computed in a similar way according to Equation \ref{Eqn:gru}. The hidden vectors for the dialog utterance $u_i^t$ can be obtained in the same procedure. Given the hidden vectors of $r_i^{k'}$ and $u_i^t$, we calculate element $m_{2,i,j}$ in the sequence hidden representation similarity matrix $\mathbf{M}_2$ by $m_{2,i,j} = \mathbf{h}_{r,i}^T \cdot \mathbf{h}_{u,j}$.
% \begin{footnotesize}
%\begin{eqnarray} \label{Eqn:matrix2_element_ij}
%\end{eqnarray}
%\end{footnotesize}
BiGRU models the neighbor context information around words from two directions and encode the text sequences into hidden vectors. Thus $\mathbf{M}_2$ matches $r_i^{k'}$ and $u_i^t$ with local sequence structures such as phrases or text segments. %\ylcomment{will update the the computation of $m_{2,i,j} $ as the bi-linear product and add one more parameter matrix here. Can also add one more non-linear transformation here.}
 
\subsubsection{\textbf{Convolution and Pooling Layers}}
% Present in a mathmatical way
The interaction matrices $\mathbf{M}_1$ and $\mathbf{M}_2$ are then fed into a CNN to learn high level matching patterns as features. CNN alternates convolution and max-pooling operations over these input channels. Let $\mathbf{z}^{(l,k)}$ denote the output feature map of the l-th layer and k-th kernel, the model will do convolution operations and max-pooling operations according to the following equations.
% in an interleaved way

\textbf{Convolution.} Let $r_w^{(l,k)} \times r_h^{(l,k)}$  denote the shape of the k-th convolution kernel in the $l$-th layer, the convolution operation can be defined as:
 \begin{footnotesize}
\begin{equation} \label{Eqn:2d_conv}
\begin{aligned}
\mathbf{z}_{i,j}^{(l+1, k)}=\sigma(\sum_{k'=0}^{K_l -1}\sum_{s=0}^{r_w^{(l,k)} - 1}\sum_{t=0}^{r_h^{(l,k)} - 1} \mathbf{w}_{s,t}^{(l+1,k)} \cdot z_{i+s, j+t}^{(l,k')} + b^{(l+1,k)}  )  \quad  \\ 
\forall l =0,2,4,6,\cdots,  
\end{aligned}
\end{equation}
\end{footnotesize}
where $\sigma$ is the activation function ReLU, and $\mathbf{w}_{s,t}^{(l+1,k)}$ and $b^{(l+1,k)}$ are the parameters of the $k$-th kernel on the $(l+1)$-th layer to be learned.  $K_l$ is the number of kernels on the $l$-th layer. 

\textbf{Max Pooling.} Let $p_w^{(l,k)} \times p_h^{(l,k)}$  denote the shape of the k-th pooling kernel in the $l$-th layer, the max pooling operation can be defined as:
 \begin{footnotesize}
\begin{equation} \label{Eqn:2d_max_pooling}
\begin{aligned}
\mathbf{z}_{i,j}^{(l+1,k)} = \max_{0\le s < p_w^{l+1, k}} \max_{0\le t < p_h^{l+1, k}} \mathbf{z}_{i+s,j+t}^{(l,k)} \quad \forall  l = 1,3,5,7, \cdots, 
\end{aligned}
\end{equation}
\end{footnotesize}
\subsubsection{\textbf{BiGRU Layer and MLP}}
% Present in a mathmatical way

%\ylcomment{Cite the ACL'17 paper on SMN and analyze the similarity and difference between DMN/DMN-PRF/DMN-KD with SMN.}

Given the output feature representation vectors learned by CNN for utterance-response pairs $(r_i^{k'}, u_i^t)$, we add another BiGRU layer to model the dependency and temporal relationship of utterances in the conversation according to Equation \ref{Eqn:gru} following the previous work \cite{DBLP:conf/acl/WuWXZL17}.  The output hidden states $\mathbf{H}_c = [\mathbf{h'}_1, \cdots, \mathbf{h'}_c]$ will be concatenated as a vector and fed into a multi-layer perceptron (MLP) to calculate the final matching score $f( \mathcal{U}_i, r_i^{k'} ) $  as
 \begin{footnotesize}
\begin{equation}
f( \mathcal{U}_i, r_i^{k'} )  = \sigma_2( \mathbf{w}_2^T \cdot \sigma_1 (\mathbf{w}_1^T\mathbf{H}_c + \mathbf{b}_1    ) + \mathbf{b}_2)
\end{equation}
\end{footnotesize}
where $\mathbf{w}_1, \mathbf{w}_2,\mathbf{b}_1,\mathbf{b}_2$ are model parameters. $\sigma_1$ and $\sigma_2$ are tanh and softmax functions respectively.

%\vspace{-0.3cm}
\begin{figure*}[th]
	\center
	\includegraphics*[viewport=0mm 0mm 310mm 95mm, scale=0.50]{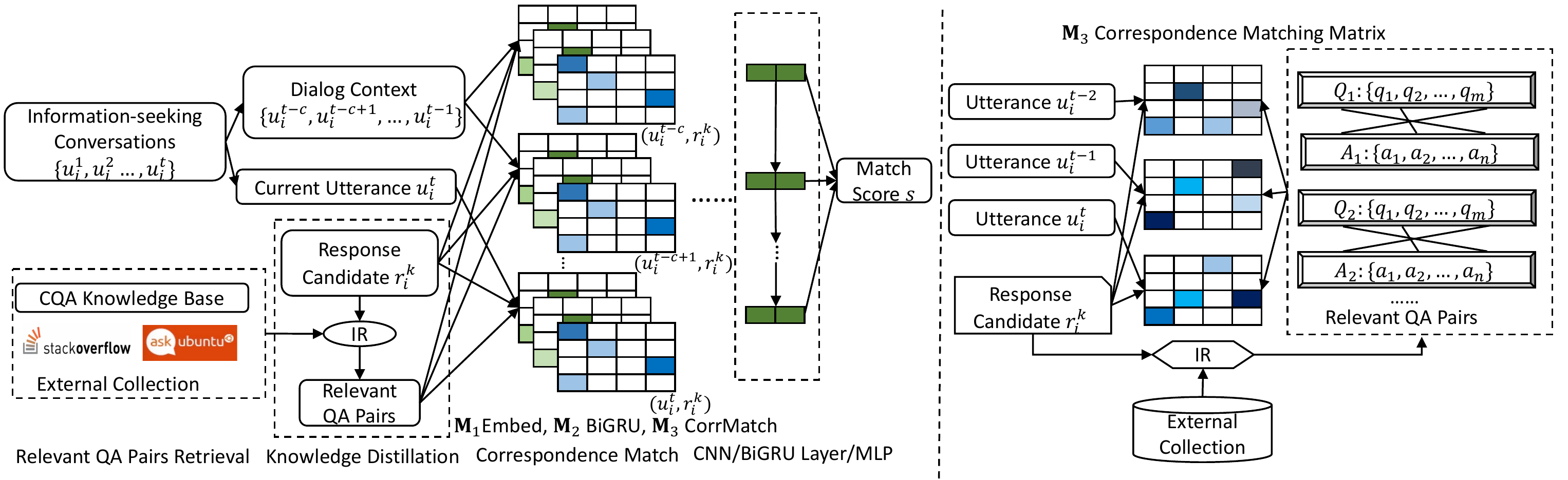}
	%\includegraphics[width=6.8in, height=3.0in]{figures/blstm-cnn}\\
	%\vspace{-0.4cm}
	\caption{The left figure shows the architecture of DMN-KD model for conversation response ranking. The input channel $\mathbf{M}_3$ denoted as blue matrices capture the correspondence matching patterns of utterance terms and response terms in relevant external QA pairs retrieved from $\mathcal{E}$. Note that we omit the details for CNN layers here to save spaces as they have been visualized in Figure \ref{fig:dmn-prf}. The right figure shows the detailed pipeline of external relevant QA pairs retrieval and QA correspondence matching knowledge distillation in DMN-KD model. }\label{fig:dmn-kd}
	\vspace{-0.4cm}
\end{figure*}

% We visualize the correspondence match matrices between $3$ dialog utterances and the response candidate $r_i^k$.

\subsubsection{\textbf{Model Training}}
% Present in a mathmatical way
For model training, we consider a pairwise ranking learning setting. The training data consists of triples $(\mathcal{U}_i, r_i^{k+}, r_i^{k-})$ where $r_i^{k+}$ and $r_i^{k-}$ denote the positive and the  negative response candidate  for dialog utterances $\mathcal{U}_i$. Let $\Theta$ denote all the parameters of our model. The pairwise ranking-based hinge loss function is defined as:
 \begin{footnotesize}
\begin{equation}
\mathcal{L}(\mathcal{D}, \mathcal{E};\Theta) = \sum_{i=1}^{I} \max(0, \epsilon - f( \mathcal{U}_i, r_i^{k+} )  +  f( \mathcal{U}_i, r_i^{k-} )  ) + \lambda ||\Theta||^2_2
\end{equation}
\end{footnotesize}
where $I$ is the total number of triples in the training data $\mathcal{D}$. $\lambda ||\Theta||^2_2$ is the regularization term where $\lambda$  denotes the regularization coefficient. $\epsilon$ denotes the margin in the hinge loss. The parameters of the deep matching network are optimized using back-propagation with \textit{Adam} algorithm~\cite{DBLP:journals/corr/KingmaB14}.  % For neural network regularization, we employ Dropout\cite{DBLP:journals/jmlr/SrivastavaHKSS14} in the model training process.

\subsection{Deep Matching Networks with QA Correspondence Knowledge Distillation}
\label{sec:method_dmn_kd}
% Present in a mathmatical way

 %\subsubsection{Knowledge Distillation from Relevant QA  Pairs}
 % Present in a mathmatical way
In addition to the DMN-PRF model presented in Section \ref{sec:method_dmn_prf}, we also propose another model for incorporating external knowledge into conversation response ranking via QA correspondence knowledge distillation, which is referred to as DMN-KD model in this paper. The architecture of DMN-KD model is presented in Figure \ref{fig:dmn-kd}. Compared with DMN-PRF, the main difference is that the CNN of DMN-KD will run on an additional input channel $\mathbf{M}_3$ denoted as blue matrices in Figure \ref{fig:dmn-kd}, which captures the correspondence matching patterns of utterance terms and response terms in relevant external QA pairs retrieved from $\mathcal{E}$. Specifically, we firstly use the response candidate $r_i^k$ as the query to retrieve a set of relevant QA pairs\footnote{Note that we want QA pairs here instead of question posts or answer posts, since we would like to extract QA term co-occurrence information with these QA pairs.}  $\mathcal{P}$. Suppose $\mathcal{P} = \{ \mathcal{Q}, \mathcal{A}\}  = \{(\mathbf{Q}_1, \mathbf{A}_1), (\mathbf{Q}_2, \mathbf{A}_2), \cdots, (\mathbf{Q}_P, \mathbf{A}_P) \}$, where $(\mathbf{Q}_p, \mathbf{A}_p)$ denotes the $p$-th QA pair. Given a response candidate $r_i^k$ and a dialog utterance $u_i^t$ in dialog $\mathcal{U}_i$, the model will compute the term co-occurrence information as the \textit{Positive Pointwise Mutual Information} (PPMI) of words of $r_i^k$ and  $u_i^t$ in retrieved QA pair set $\{ \mathcal{Q}, \mathcal{A}\} $. Let $[w_{r,1}, w_{r,2}, \cdots, w_{r, l_r}]$ and $[w_{u,1}, w_{u,2}, \cdots, w_{u, l_u}]$ denote the word sequence in $r_i^k$ and  $u_i^t$. We construct a QA term correspondence matching matrix $\mathbf{M}_3$ as the third input channel of CNN for $r_i^k$ and  $u_i^t$ with the PPMI statistics from $\{ \mathcal{Q}, \mathcal{A}\} $. More specifically, $\forall i, j$, the element $m_{3,i,j}$ in  $\mathbf{M}_3$ is computed as 

\vspace{-0.1in}
 \begin{footnotesize}
\begin{eqnarray}\label{Eqn:kd_co_occurrence_matrix}
m_{3,i,j} &=& PPMI(w_{r,i}, w_{u,j}|\{\mathcal{Q}, \mathcal{A}\})    \\
&=& \max(0, \log \frac{ \sum_{p'=1}^P p(w_{r,i} \in \mathbf{A}_{p'}, w_{u,j} \in \mathbf{Q}_{p'}|\mathbf{Q}_{p'}, \mathbf{A}_{p'})  }{p(w_{r,i}|\mathcal{A}) \cdot p(w_{u,j}|\mathcal{Q})}) \nonumber
\end{eqnarray}
\vspace{-0.1in}
\end{footnotesize}

%\begin{footnotesize}
%	\begin{eqnarray}
%	&&p(z_{c} = z, e_{c} = e|\mathbf{Z}_{\neg_{c}}, \mathbf{W}, \mathbf{E}_{\neg_{c}},\mathbf{V}, \mathbf{T}, \Theta) \nonumber \\
%	&\propto& \frac{p(\mathbf{Z}, \mathbf{W}, \mathbf{E},\mathbf{V}, \mathbf{T} | \Theta)}
%	{p(\mathbf{Z}_{\neg_{c}}, \mathbf{W}, \mathbf{E}_{\neg_{c}},\mathbf{V}, \mathbf{T} | \Theta)} \nonumber \\
%	&=& \frac{\Delta(C_{u}^\mathbf{k} + \alpha)}{\Delta(C_{u,\neg_{c}}^\mathbf{k} + \alpha)} \cdot
%	\frac{\Delta(C_{z}^\mathbf{w} + \gamma)}{\Delta(C_{z,\neg_{c}}^\mathbf{w} + \gamma)} \cdot
%	\frac{\Delta(C_{z}^\mathbf{t} + \eta)}{\Delta(C_{z,\neg_{c}}^\mathbf{t} + \eta)} \nonumber \\
%	&&   \cdot \frac{\Delta(C_{z,u}^\mathbf{e} + \beta)}{\Delta(C_{z,u,\neg_{c}}^\mathbf{e} + \beta)} \cdot \mathcal{N}(v_c | \mu_e, \Sigma_e)  \nonumber \\
%	&=& \frac{C_{u,\neg_c}^{z} + \alpha}{\sum_{k=1}^{K}C_{u,\neg_c}^{k} + K \alpha} \cdot
%	\frac{\prod_{w=1}^{V} \prod_{i=1}^{n_c^w} (C_{z,\neg_c}^{w} + \gamma + i - 1)}
%	{\prod_{j=1}^{n_c^{\mathbf{w}}} \sum_{w=1}^V (C_{z,\neg_c}^{w} + V\gamma + j - 1)} \nonumber \\
%	&&  \cdot \frac{\prod_{t=1}^{T} \prod_{p=1}^{n_c^t} (C_{z,\neg_c}^{t} + \eta + p - 1)}
%	{\prod_{q=1}^{n_c^{\mathbf{t}}} \sum_{t=1}^T (C_{z,\neg_c}^{t} + T\eta + q - 1)} \nonumber \\
%	&&  \cdot \frac{C_{z,u,\neg_c}^{e} + \beta}{\sum_{e=1}^{E}C_{z,u,\neg_c}^{e} + E \beta} \cdot
%	\mathcal{N}(v_c | \mu_e, \Sigma_e),  \label{eqn_gibbsRuleRes}
%	\end{eqnarray}
%\end{footnotesize}
where $w_{r,i}$ and $w_{u,j}$ denote the $i$-th word in the response candidate and $j$-th word in the dialog utterance.  The intuition is that the PPMI between $w_{r,i}$ and $w_{u,j}$ in the top retrieved relevant QA pair set $\{\mathcal{Q}, \mathcal{A}  \}$ could encode the correspondence matching patterns between $w_{r,i}$ and $w_{u,j}$ in external relevant QA pairs . Thus $\mathbf{M}_3$ is the extracted QA correspondence  knowledge from the external collection $\mathcal{E}$ for $r_i^k$ and $u_i^t$. These correspondence matching knowledge capture relationships such as \textit{``(Problem Descriptions, Solutions)'', ``(Symptoms, Causes)'', ``(Information Request, Answers)''}, etc. in  the top ranked relevant QA pair set $\{\mathcal{Q}, \mathcal{A}  \}$. They will help the model better discriminate a good response candidate from a bad response candidate given the dialog context utterances. To compute the co-occurrence count between $w_{r,i}$ and $w_{u,j}$, we count all word co-occurrences considering $\mathbf{A}_p$ and $\mathbf{Q}_p$ as bag-of-words as we found this setting is more effective in experiments.

\section{Experiments}
\label{sec:exps}
 % Present it in a mathematical way
 %\ylcomment{Until 01/21 18:57 pm , removing duplicated or verbose content/sentences to here.}
 \subsection{Data Set Description}
 \label{sec:data_desc}

 We evaluated our method with three data sets: Ubuntu Dialog Corpus (UDC), MSDialog, and AliMe  data consisting of a set of %internal commercial Chinese 
 customer service conversations in Chinese from Alibaba.
 \begin{table}[]
	\footnotesize
	%\scriptsize
	\centering
	\caption{The statistics of experimental datasets, where C denotes context and R denotes response. \# Cand. per C denotes the number of candidate responses per context.}
	\vspace{-0.1in}
	\label{tab:exp_data_stat_train_valid_test}
	\begin{tabular}{l | p{0.35cm} p{0.35cm} p{0.35cm} | p{0.35cm} p{0.35cm} p{0.35cm} | p{0.3cm} p{0.3cm} p{0.3cm}}
		\hline \hline
		Data                               & \multicolumn{3}{c|}{UDC}      & \multicolumn{3}{c|}{MSDialog} & \multicolumn{3}{c}{AliMe} \\ \hline
		Items                              & Train     & Valid   & Test    & Train     & Valid   & Test    & Train    & Valid  & Test   \\ \hline
		\# C-R pairs          & 1000k & 500k & 500k & 173k   & 37k  & 35k  & 51k   & 6k  & 6k  \\ \hline
		\# Cand. per C          & 2         & 10      & 10      & 10        & 10      & 10      & 15       & 15     & 15     \\ \hline
		\# + Cand. per C & 1         & 1       & 1       & 1         & 1       & 1       & 2.9      & 2.8    & 2.9    \\ \hline
		Min \# turns per C           & 1         & 2       & 1       & 2         & 2       & 2       & 2        & 2      & 2      \\ \hline
		Max \# turns per C           & 19        & 19      & 19      & 11        & 11      & 11      & 3        & 3      & 3      \\ \hline
		Avg \# turns per C           & 10.1      & 10.1    & 10.1    & 5.0       & 4.9     & 4.4     & 2.4      & 2.1    & 2.2    \\ \hline
		Avg \# words per C           & 116     & 116   & 116   & 271     & 263   & 227   & 38     & 35   & 34   \\ \hline
		Avg \# words per R          & 22.1      & 22.1    & 22.1    & 66.7      & 67.6    & 66.8    & 4.9      & 4.7    & 4.6    \\ \hline \hline
	\end{tabular}
\end{table}

\subsubsection{\textbf{Ubuntu Dialog Corpus}} The Ubuntu Dialog Corpus (UDC) \cite{DBLP:journals/corr/LowePSP15} contains multi-turn technical support conversation data collected from the chat logs of the Freenode Internet Relay Chat (IRC) network. We used the data copy shared by Xu et al.\cite{DBLP:journals/corr/XuLWSW16}, in which numbers, urls and paths are replaced by special placeholders. It is also used in several previous related works \cite{DBLP:conf/acl/WuWXZL17}\footnote{The data can be downloaded from \url{https://www.dropbox.com/s/2fdn26rj6h9bpvl/ubuntu\%20data.zip?dl=0}}. It consists of $1$ million context-response pairs for training, $0.5$ million pairs for validation and  $0.5$ million pairs for testing. The statistics of this data is shown in Table \ref{tab:exp_data_stat_train_valid_test}. The positive response candidates in this data come form the true responses by human and negative response candidates are randomly sampled. %In average, there are around $10$ turns in the dialog context and $22$ words in each utterance. 

\subsubsection{\textbf{MSDialog}}
 % Data preprocess, response candidate sampling, data spliting and statistics for train/valid/test data sets of the MSDialog data
  In addition to UDC, we also crawled another technical support conversation data from the Microsoft Answer community, which is a QA forum on topics about a variety of Microsoft products. We firstly crawled $35,536$ dialogs about $76$ different categories of Microsoft products including ``Windows'', ``IE'', ``Office'', ``Skype'', ``Surface'', ``Xbox'', etc. \footnote{Note that some categories are more fine-grained, such as``Outlook\_Calendar'', ``Outlook\_Contacts'', ``Outlook\_Email'', ``Outlook\_Messaging'', etc.} Then we filtered dialogs whose number of turns are out of the range $[3,99]$. After that we split the data into training/validation/testing partitions by time. Specifically, the training data contains $25,019$ dialogs from ``2005-11-12'' to ``2017-08-20''. The validation data contains $4,654$ dialogs from ``2017-08-21'' to ``2017-09-20''. The testing data contains $5,064$ dialogs from ``2017-09-21'' to ``2017-10-04''. 
  
  The next step is to generate the dialog context and response candidates. For each dialog, we assigned ``User'' label to the first participant who proposed the question leading to this information-seeking conversation, and ``Agent'' label to the other participants who provided responses. The ``Agent'' in our data could be Microsoft customer service staff, a Microsoft MVP (Most Valuable Professional) or a user from the Microsoft Answer community. Then for each utterance by the ``User'' $u_i^t$ \footnote{We consider the utterances by the user except the first utterance, since there is no associated dialog context with it. }, we collected the previous $c$ utterances as the dialog context, where $c = \min(t-1,10)$ and $t-1$ is the total number of utterances before $u_i^t$. The true response by the ``Agent'' becomes the positive response candidate. For the negative response candidates, we adopted negative sampling to construct them following previous work \cite{DBLP:conf/aaai/WanLGXPC16,DBLP:journals/corr/LowePSP15,DBLP:conf/acl/WuWXZL17}. For each dialog context, we firstly used the true response as the query to retrieve the top $1,000$ results from the whole response set of agents with BM25. Then we randomly sampled $9$ responses from them to construct the negative response candidates. The statistics of MSDialog data is presented in Table \ref{tab:exp_data_stat_train_valid_test}. For data preprocessing, we performed tokenization and punctuation removal. Then we removed stop words and performed word stemming. For neural models, we also removed words that appear less than $5$ times in the whole corpus. 
  %  The used different QA post collections as source of external knowledge (e.g. Stack Overflow data for MSDialog, AskUbuntu for UDC) have been presented in Table \ref{tab:data_summery}. 
  % with NLTK \footnote{\url{https://www.nltk.org/}} toolkit. 
  
\subsubsection{\textbf{AliMe Data}}
We collected the chat logs between customers and a chatbot AliMe from ``2017-10-01'' to ``2017-10-20'' in Alibaba. %We filter out single turn conversation and focus our study on conversations within 3 turns~\footnote{The majority (around $85\%$) of conversations in the dataset are within 3 turns.}. %We randomly sampled 4200 sessions, and for each session, we concatenate all the utterances to 
The chatbot is built based on a question-to-question matching system~\footnote{
Interested readers can access AliMe Assist through the Taobao App, or the web version via \url{https://consumerservice.taobao.com/online-help}}~\cite{alime-demo}, where for each query, it finds the most similar candidate question in a QA database and return its answer as the reply. It indexes all the questions in our QA database using Lucence\footnote{\url{https://lucene.apache.org/core/}}. For each given query, it uses TF-IDF ranking algorithm to call back candidates. %top-K~\footnote{We set K=15.} most similar candidate questions as candidate `responses'~\footnote{A `response' here is a question in our QA system.}. 
To form our data set, we concatenated utterances within three turns~\footnote{The majority (around $85\%$) of conversations in the data set are within 3 turns.} to form a query, and used the chatbot system to call back top-K ~\footnote{We set K=15.} most similar candidate questions as candidate ``responses''.~\footnote{A ``response'' here is a question in our system.} We then asked a business analyst to annotate the candidate responses, where a ``response'' is labeled as positive if it matches the query, otherwise negative. In all, we have annotated 63,000 context-response pairs, where we use 51,000 as training, 6,000 for testing, and 6,000 for validation shown in Table \ref{tab:exp_data_stat_train_valid_test}. Note that we have included human evaluation in AliMe data. Furthermore, if the confidence score of answering a given user query is low, the system will prompt three top related questions for users to choose. We collected such user click logs as our external data, where we treat the clicked question as positive and the others as negative. We collected 510,000 clicked questions with answers from the click logs in total as the source of external knowledge. %\ylcomment{ToDo: Double check the number $51,000$ and $510,000$ here. Should the ``context-response pairs from the click logs'' here be changed to ``retrieved QA pairs from the click logs ?''}

\begin{table*}[]
	\footnotesize
	\centering
	\caption{Comparison of different models over Ubuntu Dialog Corpus (UDC), MSDialog, and AliMe data sets. 
		Numbers in bold font mean the result is better compared with the best baseline. 
		$\ddagger$ means statistically significant difference over the best baseline with $p < 0.05$ measured by the Student's paired t-test.} % \ylcomment{needs score files of SMN on MS\_V2 for sig test}
	\vspace{-0.1in}
	\label{tab:exp_res_udc_ms_ec}
	\begin{tabular}{l|l l l l | l l l l | l l l l}
		\hline \hline
		Data     & \multicolumn{4}{c|}{UDC}                & \multicolumn{4}{c|}{MSDialog}           & \multicolumn{4}{c}{AliMe}              \\ \hline
		Methods  & MAP    & Recall@5 & Recall@1 & Recall@2 & MAP    & Recall@5 & Recall@1 & Recall@2 & MAP    & Recall@5 & Recall@1 & Recall@2 \\ \hline \hline
		BM25     & 0.6504 & 0.8206   & 0.5138   & 0.6439   & 0.4387 & 0.6329   & 0.2626   & 0.3933   & 0.6392 & 0.6407   & 0.2371   & 0.4204   \\ \hline
		BM25-PRF & 0.6620 & 0.8292   & 0.5289   & 0.6554   & 0.4419 & 0.6423   & 0.2652   & 0.3970   & 0.6412 & 0.6510 & 0.2454         &  0.4209        \\ \hline
		ARC-II   & 0.6855 & 0.8978   & 0.5350   & 0.6959   & 0.5398 & 0.8662   & 0.3189   & 0.5413   & 0.7306 & 0.6595   & 0.2236   & 0.3671   \\ \hline
		MV-LSTM  & 0.6611 & 0.8936   & 0.4973   & 0.6733   & 0.5059 & 0.8516   & 0.2768   & 0.5000   & 0.7734 & 0.7017   & 0.2480   & 0.4105   \\ \hline
		DRMM     & 0.6749 & 0.8776   & 0.5287   & 0.6773   & 0.5704 & 0.9003   & 0.3507   & 0.5854   & 0.7165 & 0.6575   & 0.2212   & 0.3616   \\ \hline
		Duet     & 0.5692 & 0.8272   & 0.4756   & 0.5592   & 0.5158 & 0.8481   & 0.2934   & 0.5046   & 0.7651 & 0.6870   & 0.2433   & 0.4088   \\ \hline
		SMN      & 0.7327 & 0.9273   & 0.5948   & 0.7523   &  0.6188 & 0.8374 & 0.4529 & 0.6195 & 0.8145 & 0.7271   & 0.2881   & 0.4680   \\ \hline
		DMN      & 0.7363 & 0.9196   & 0.6056   & 0.7509   & 0.6415 & 0.9155   & 0.4521   & 0.6673   & 0.7833 & 0.7629   & 0.3568   & 0.5012   \\ \hline \hline
		DMN-KD   & \textbf{0.7655}$^\ddagger$ & \textbf{0.9351}$^\ddagger$   & \textbf{0.6443}$^\ddagger$   & \textbf{0.7841}$^\ddagger$   & \textbf{0.6728}$^\ddagger$ & \textbf{0.9304}$^\ddagger$   & \textbf{0.4908}$^\ddagger$   & \textbf{0.7089}$^\ddagger$   & \textbf{0.8323} & \textbf{0.7631}   & \textbf{0.3596}$^\ddagger$   & \textbf{0.5122}$^\ddagger$  \\ \hline
		DMN-PRF  & \textbf{0.7719}$^\ddagger$ & \textbf{0.9343}$^\ddagger$   & \textbf{0.6552}$^\ddagger$   & \textbf{0.7893}$^\ddagger$   & \textbf{0.6792}$^\ddagger$ & \textbf{0.9356}$^\ddagger$   & \textbf{0.5021}$^\ddagger$   & \textbf{0.7122}$^\ddagger$   & \textbf{0.8435}$^\ddagger$ & \textbf{0.7701} $^\ddagger$  & \textbf{0.3601} $^\ddagger$  & \textbf{0.5323} $^\ddagger$  \\ \hline \hline
	\end{tabular}
\end{table*}

\subsection{Experimental Setup}

\subsubsection{\textbf{Baselines.}}
We consider different types of baselines for comparison, including traditional retrieval models, deep text matching models and the state-of-the-art multi-turn conversation response ranking method as the following:

%\begin{itemize}
 \textbf{BM25.} This method uses the dialog context as the query to retrieve response candidates for response selection. We consider BM25 model \cite{Robertson:1994:SEA:188490.188561} as the retrieval model.

 \textbf{ARC-II.} ARC-II is an interaction focused deep text matching architectures proposed by Hu et al. \cite{DBLP:conf/nips/HuLLC14}, which is built directly on the interaction matrix between the dialog context and response candidates. A CNN is running on the interaction matrix to learn the matching representation score.
 %It could let the two text sequences meet before their own high-level representations mature, while still retaining the space for the individual development of abstraction of each text sequence. 
 % It is essentially the Siamese CNN architecture. 

% \textbf{DSSM.} DSSM \cite{DBLP:conf/cikm/HuangHGDAH13} is neural text matching model proposed from Web search and trained with click-through data. It consists of a word hashing layer, two non-linear feed forward hidden layers and an output layer.
%
% \textbf{CDSSM.} CDSSM \cite{Shen:2014:LSR:2567948.2577348} is a similar neural text matching model with DSSM. It uses CNN to learn low dimensional semantic vectors for the dialog context and response candidate. 
% 
% \ylcomment{To be updated. Since DSSM/CDSSM need large click-through data for effective training, we need to directly use the released models Sent2Vec on MS Web (trained on large click-through dataset) on our test data. Training DSSM/CDSSM on smaller data sets are likely to overfit the data. Or we can also delete these two baselines.}
 
 \textbf{MV-LSTM.} MV-LSTM \cite{DBLP:conf/aaai/WanLGXPC16} is a neural text matching model that matches two sequences with multiple positional representations learned by a Bi-LSTM layer. 
 %Then it models the interactions between a pair of text sequences from different positions. The matching score is finally produced by aggregating interactions through k-Max pooling and a multi-layer perceptron.

% \textbf{MatchPyramid.} MatchPyramid \cite{DBLP:conf/aaai/PangLGXWC16} firstly constructs a word level interaction similarity matrix of two text sequences to capture the basic word level matching signals and then applies a CNN on this matrix to learn representations of matching patterns. 
 
 %Dynamic pooling strategy is used to deal with the text length variability.
 \textbf{DRMM.} DRMM \cite{Guo:2016:DRM:2983323.2983769} is a deep relevance matching model for ad-hoc retrieval. We implemented a variant of DRMM for short text matching. Specifically, the matching histogram is replaced by a top-k max pooling layer and the remaining part is the same with the original model.
 
 \textbf{Duet.} Duet \cite{Mitra:2017:LMU:3038912.3052579} is the state-of-the-art deep text matching model that jointly learns local lexical matching and global semantic matching between the two text sequences. 

% \textbf{DL2R.} Deep Learning to Respond (DL2R) \cite{DBLP:conf/sigir/YanSW16} is a multi-turn conversation response ranking method, which enhances the current utterance by adding its contexts. Then the model uses a BiLSTM layer to propagate information across words; a CNN layer further captures patterns of adjacent words. Then a matching layer combines the information in each individual sentence.

 \textbf{SMN.} Sequential Matching Network (SMN) \cite{DBLP:conf/acl/WuWXZL17} is the state-of-the-art deep neural architecture for multi-turn conversation response selection. It matches a response candidate with each utterance in the context on multiple levels of granularity and then adopts a CNN network to distill matching features. We used the TensorFlow \footnote{\url{https://www.tensorflow.org/}} implementation of SMN shared by authors \cite{DBLP:conf/acl/WuWXZL17} \footnote{The reported SMN results with the code from authors are on the raw data sets of UDC and MSDialog without any over sampling of negative training data.}.
 % The final matching score is aggregated from the hidden states of a RNN layer on the output of the CNN network. 
%\end{itemize}

We also consider a degenerated version of our model, denoted as \textbf{DMN}, where we do not incorporate external knowledge via pseudo-relevance feedback or QA correspondence knowledge distillation. Finally, we consider a baseline \textbf{BM25-PRF}, where we incorporate external knowledge into BM25 by matching conversation context with the expanded responses as in Section \ref{sec:response_expansion} using BM25 model.

\subsubsection{\textbf{Evaluation Methodology}.}
For the evaluation metrics, we adopted mean average precision (MAP), Recall@1, Recall@2, and Recall@5 following previous related works \cite{DBLP:conf/acl/WuWXZL17,DBLP:journals/corr/LowePSP15}. For UDC and MSDialog,  MAP is equivalent to the mean reciprocal rank (MRR) since there is only one positive response candidate per dialog context. For AliMe data, each dialog context could have more than one positive response candidates.

\subsubsection{\textbf{ Parameter Settings}.} 
All models were implemented with TensorFlow and MatchZoo\footnote{\url{https://github.com/faneshion/MatchZoo}} toolkit. Hyper-parameters are tuned with the validation data. For the hyper-parameter settings of DMN-KD and DMN-PRF models, we set the window size of the convolution and pooling kernels as $(3,3)$. The number of convolution kernels is $8$ for UDC and $2$ for MSDialog. The dimension of the hidden states of BiGRU layer is set as $200$ for UDC  and $100$ for MSDialog . The dropout rate is set as $0.3$ for UDC  and $0.6$ for MSDialog . All models are trained on a single Nvidia Titan X GPU by stochastic gradient descent with Adam\cite{DBLP:journals/corr/KingmaB14} algorithm. The initial learning rate is $0.001$. The parameters of Adam, $\beta_1$ and $\beta_2$ are $0.9$ and $0.999$ respectively. The batch size is $200$ for UDC and $50$ for MSDialog. The maximum utterance length is $50$ for UDC and $90$ for MSDialog. The maximum conversation context length is set as $10$ following previous work \cite{DBLP:conf/acl/WuWXZL17}. We padded zeros if the number of utterances in a context is less than $10$. Otherwise the most recent $10$ utterances will be kept. For DMN-PRF, we retrieved top $10$ QA posts and extracted $10$ terms as response expansion terms. For DMN-KD, we retrieved top $10$ question posts with accepted answers. For the word embeddings used in our experiments, we trained word embeddings with the Word2Vec tool \cite{DBLP:conf/nips/MikolovSCCD13} with the Skip-gram model using our training data. The max skip length between words and the number of negative examples is set as $5$ and $10$ respectively. The dimension of word vectors is $200$. Word embeddings will be initialized by these pre-trained word vectors and updated during the training process.

% The number of convolution/pooling layers is set as $1$ as adding more layers would lead to over-fitting on our training data. 

%\ylcomment{Revise and double check the camera ready version to here until 18:16pm on 04/25/2018.}
%\subsubsection{\textbf{ Word Embeddings}.}
\subsection{Evaluation Results}
\subsubsection{\textbf{Performance Comparison on UDC and MSDialog}}

We present evaluation results over different methods on UDC and MSDialog in Table \ref{tab:exp_res_udc_ms_ec}. We summarize our observations as follows: (1) DMN-PRF model outperforms all the baseline methods including traditional retrieval models, deep text matching models and the state-of-the-art SMN model for response ranking on both conversation datasets. The results demonstrate that candidate response expansion with pseudo-relevance feedback could improve the ranking performance of responses in conversations. The main difference between DMN-PRF model and SMN model is the information extracted from retrieved feedback QA posts as external knowledge. This indicates the importance of modeling external knowledge with pseudo-relevant feedback beyond the dialog context for response selection. (2) DMN-KD model also outperforms all the baseline methods on MSDialog and UDC. These results show that the extracted QA correspondence matching knowledge could help the model select better responses. Comparing DMN-KD and DMN-PRF, their performances are very close. (3) If we compare the performances of DMN-PRF, DMN-KD with the degenerated model DMN, we can see that incorporating external knowledge via both pseudo-relevance feedback and QA correspondence knowledge distillation could improve the performance of the deep neural networks for response ranking with large margins. For example, the improvement of DMN-PRF against DMN on UDC is $4.83\%$ for MAP, $1.60\%$ for Recall@5, $8.19\%$ for Recall@1, $5.11\%$ for Recall@2 respectively. The differences are statistically significant with $p < 0.05$ measured by the Student's paired t-test. %These results justify that the performance of response selection in information-seeking conversations could be improved by incorporating external knowledge. Pseudo relevance feedback and QA knowledge correspondence distillation from relevant external QA pairs are effective methods to achieve this goal. ; the improvement of DMN-PRF against DMN on MSDialog is $\%17.16$ for MAP, $\%7.57$ for Recall@5, $\%31.71$ for Recall@1, $\%22.20$ for Recall@2 respectively. 

\begin{table*}[]
	\footnotesize
	\centering
	\caption{Evaluation results of model ablation. ``TB\ref{tab:exp_res_udc_ms_ec}'' means the setting is the same with the results in Table \ref{tab:exp_res_udc_ms_ec}. For DMN-KD, the model is the same with DMN if we remove M3. Numbers in bold font mean the result is better compared with other settings.}
	\vspace{-0.1in}
	\label{tab:model_ablation_results}
	\begin{tabular}{l|l|l|l|l|l|l|l|l|l}
		\hline \hline
		& Data            & \multicolumn{4}{c|}{UDC}                                              & \multicolumn{4}{c}{MSDialog}                                         \\ \hline
		Model                    & Change          & MAP             & Recall@5        & Recall@1        & Recall@2        & MAP             & Recall@5        & Recall@1        & Recall@2        \\ \hline \hline
		\multirow{5}{*}{DMN-PRF} & Only M1         & 0.7599          & 0.9294          & 0.6385          & 0.7761          & 0.5632          & 0.8509          & 0.3654          & 0.5579          \\ \cline{2-10} 
		& Only M2         & 0.7253          & 0.9271          & 0.5836          & 0.7440          & 0.4996          & 0.8584          & 0.2595          & 0.5021          \\ \cline{2-10} 
		& Inter-Dot (TB5) & \textbf{0.7719} & \textbf{0.9343} & \textbf{0.6552} & \textbf{0.7893} & \textbf{0.6792} & \textbf{0.9356} & \textbf{0.5021} & \textbf{0.7122} \\ \cline{2-10} 
		& Inter-Cosine    & 0.7507          & 0.9260          & 0.6248          & 0.7675          & 0.6729          & 0.9356          & 0.4944          & 0.7027          \\ \cline{2-10} 
		& Inter-Bilinear  & 0.7228          & 0.9199          & 0.5829          & 0.7401          & 0.4923          & 0.8421          & 0.2647          & 0.4744          \\ \hline
		\multirow{9}{*}{DMN-KD}  & Only M1         & 0.7449          & 0.9247          & 0.6167          & 0.7612          & 0.5776          & 0.8673          & 0.3805          & 0.5779          \\ \cline{2-10} 
		& Only M2         & 0.7052          & 0.9203          & 0.5538          & 0.7260          & 0.5100          & 0.8613          & 0.2794          & 0.5011          \\ \cline{2-10} 
		& Only M3         & 0.3887          & 0.6017          & 0.2015          & 0.3268          & 0.3699          & 0.6650          & 0.1585          & 0.2957          \\ \cline{2-10} 
		& M1+M2 (DMN)     & 0.7363          & 0.9196          & 0.6056          & 0.7509          & 0.6415          & 0.9155          & 0.4521          & 0.6673          \\ \cline{2-10} 
		& M1+M3           & 0.7442          & 0.9251          & 0.6149          & 0.7612          & 0.6134          & 0.8860          & 0.4224          & 0.6266          \\ \cline{2-10} 
		& M2+M3           & 0.7077          & 0.9198          & 0.5586          & 0.7263          & 0.5141          & 0.8659          & 0.2885          & 0.5069          \\ \cline{2-10} 
		& Inter-Dot (TB5) & \textbf{0.7655} & \textbf{0.9351} & \textbf{0.6443} & \textbf{0.7841} & 0.6728          & \textbf{0.9304} & 0.4908          & 0.7089          \\ \cline{2-10} 
		& Inter-Cosine    & 0.7156          & 0.9121          & 0.5770          & 0.7268          & \textbf{0.6916} & 0.9249          & \textbf{0.5241} & \textbf{0.7249} \\ \cline{2-10} 
		& Inter-Bilinear  & 0.7061          & 0.9135          & 0.5590          & 0.7225          & 0.4936          & 0.8224          & 0.2679          & 0.4814          \\ \hline \hline
	\end{tabular}
\end{table*}

\subsubsection{\textbf{Performance Comparison on AliMe Data}}

We further compare our models with the competing methods on AliMe data in Table~\ref{tab:exp_res_udc_ms_ec}. We find that: (1) our DMN model has comparable results in terms of MAP when compared with SMN, but has better Recall; (2) DMN-KD shows comparable or better results than all the baseline methods; (3) DMN-PRF significantly outperforms other competing baselines which shows the effectiveness of adding external pseudo-relevance feedback to the task; (4) both DMN-PRF and DMN-KD show better results than DMN, which demonstrates the importance of incorporating external knowledge via both pseudo-relevance feedback and QA correspondence knowledge distillation. 
%our model results ... \ylcomment{To be added by MH.}

 %\vspace{-0.3cm}
\begin{figure}[th]
	\center
	\includegraphics*[viewport=0mm 0mm 160mm 120mm, scale=0.4]{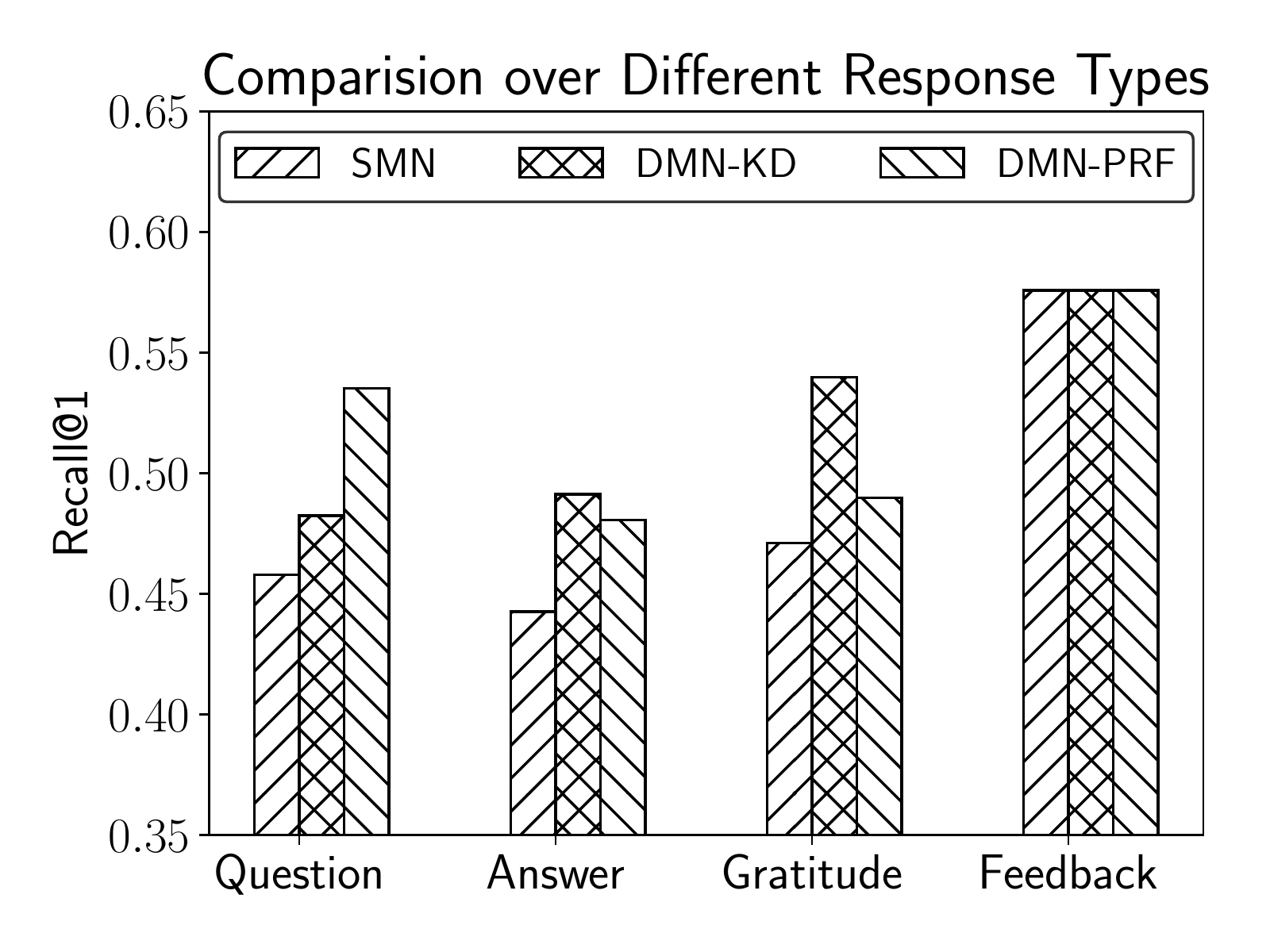}
	\vspace{-0.4cm}
	\caption{Performance comparison over different response types on MSDialog data.}\label{fig:result_diff_response_type}
	%\vspace{-0.4cm}
\end{figure}

\begin{figure*}
	\centering
	\begin{subfigure}[b]{0.245\textwidth}
		\includegraphics[width=\textwidth]{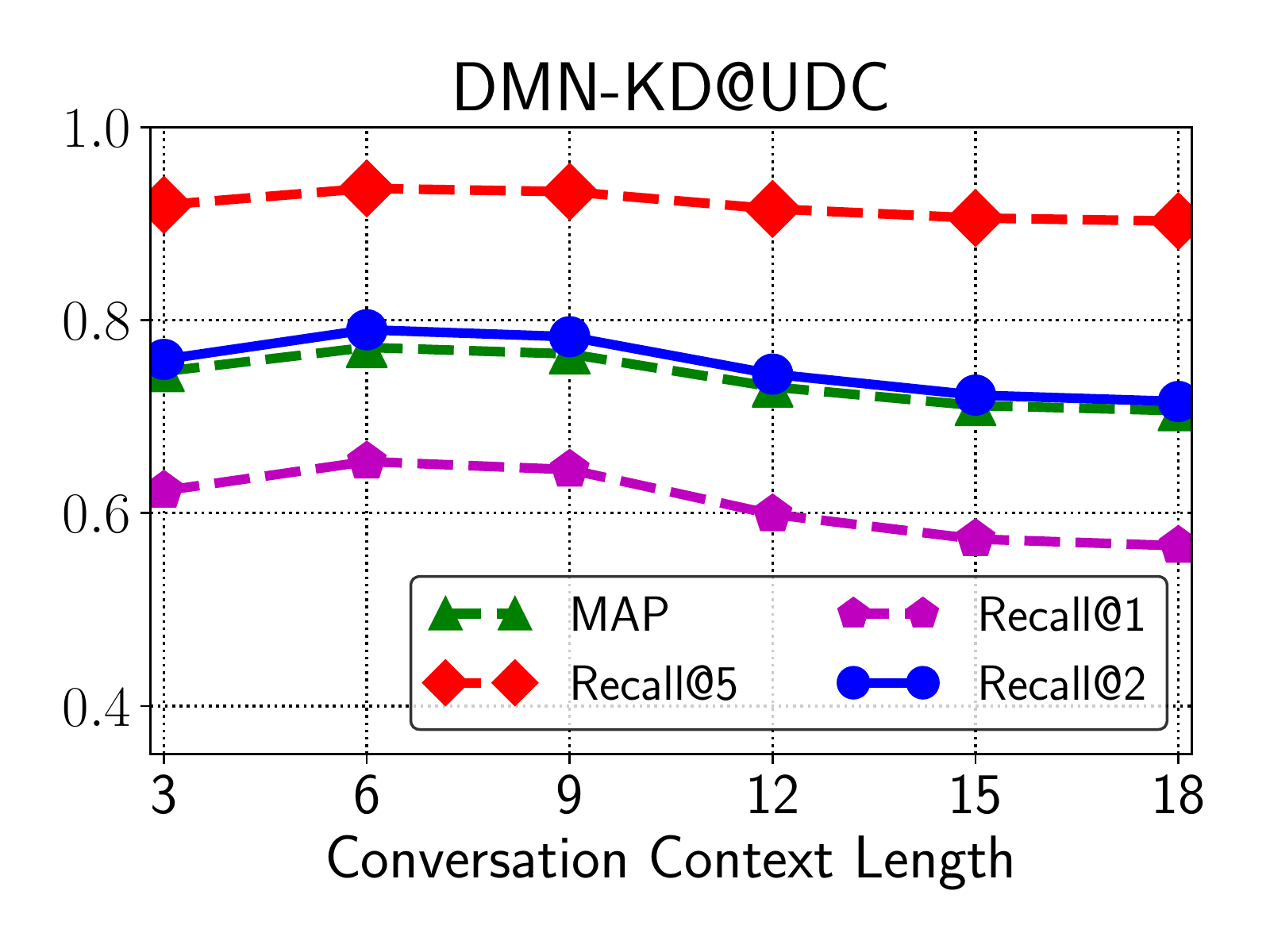}
		\label{fig:convlen-udc-dmnkd}
	\end{subfigure}
	\hspace{-0.1in}
	~ %add desired spacing between images, e. g. ~, \quad, \qquad, \hfill etc. 
	\begin{subfigure}[b]{0.245\textwidth}
		\includegraphics[width=\textwidth]{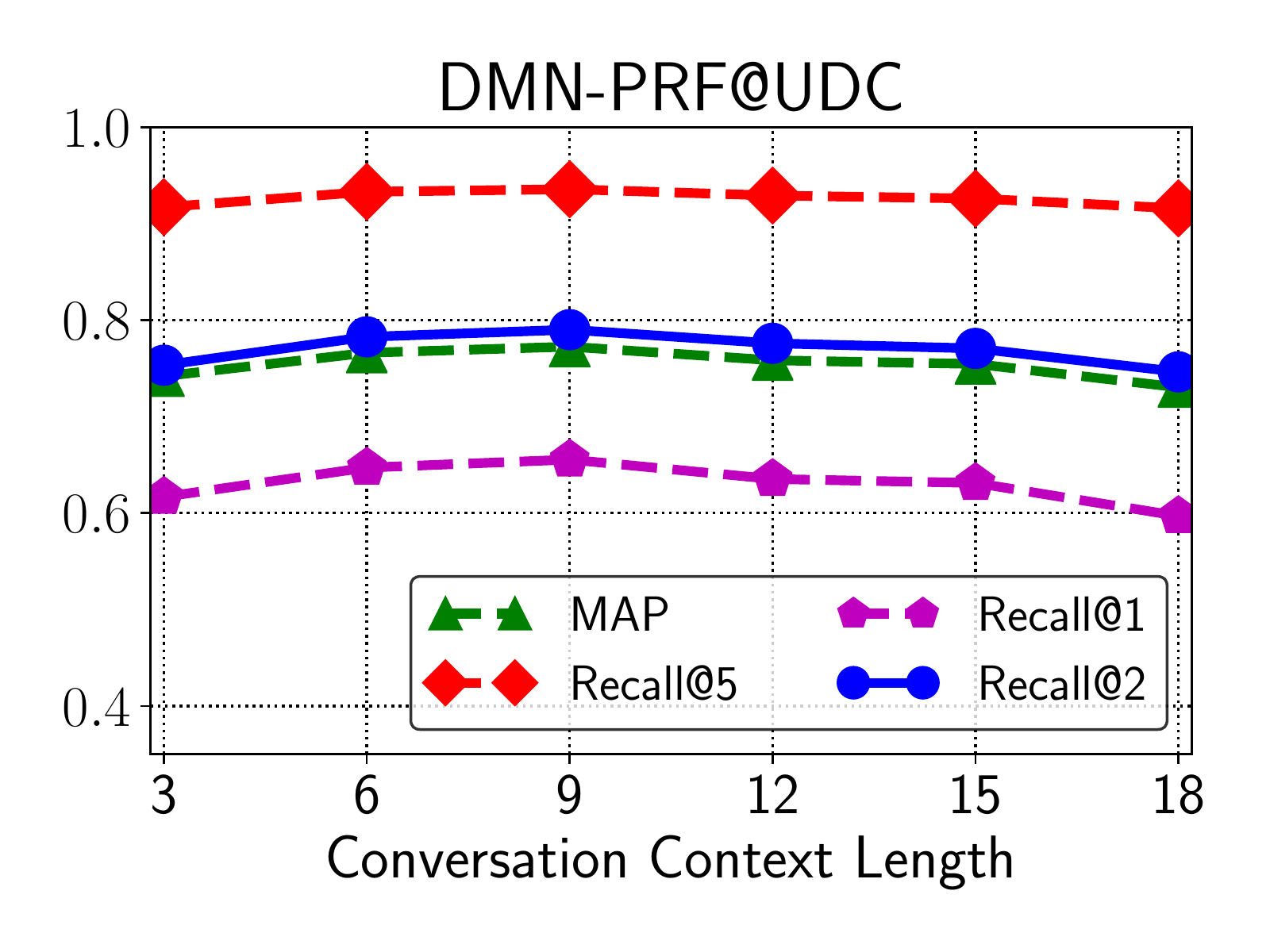}
		\label{fig:convlen-udc-dmnprf}
	\end{subfigure}
	\hspace{-0.1in}
	~ %add desired spacing between images, e. g. ~, \quad, \qquad, \hfill etc. 
	\begin{subfigure}[b]{0.245\textwidth}
		\includegraphics[width=\textwidth]{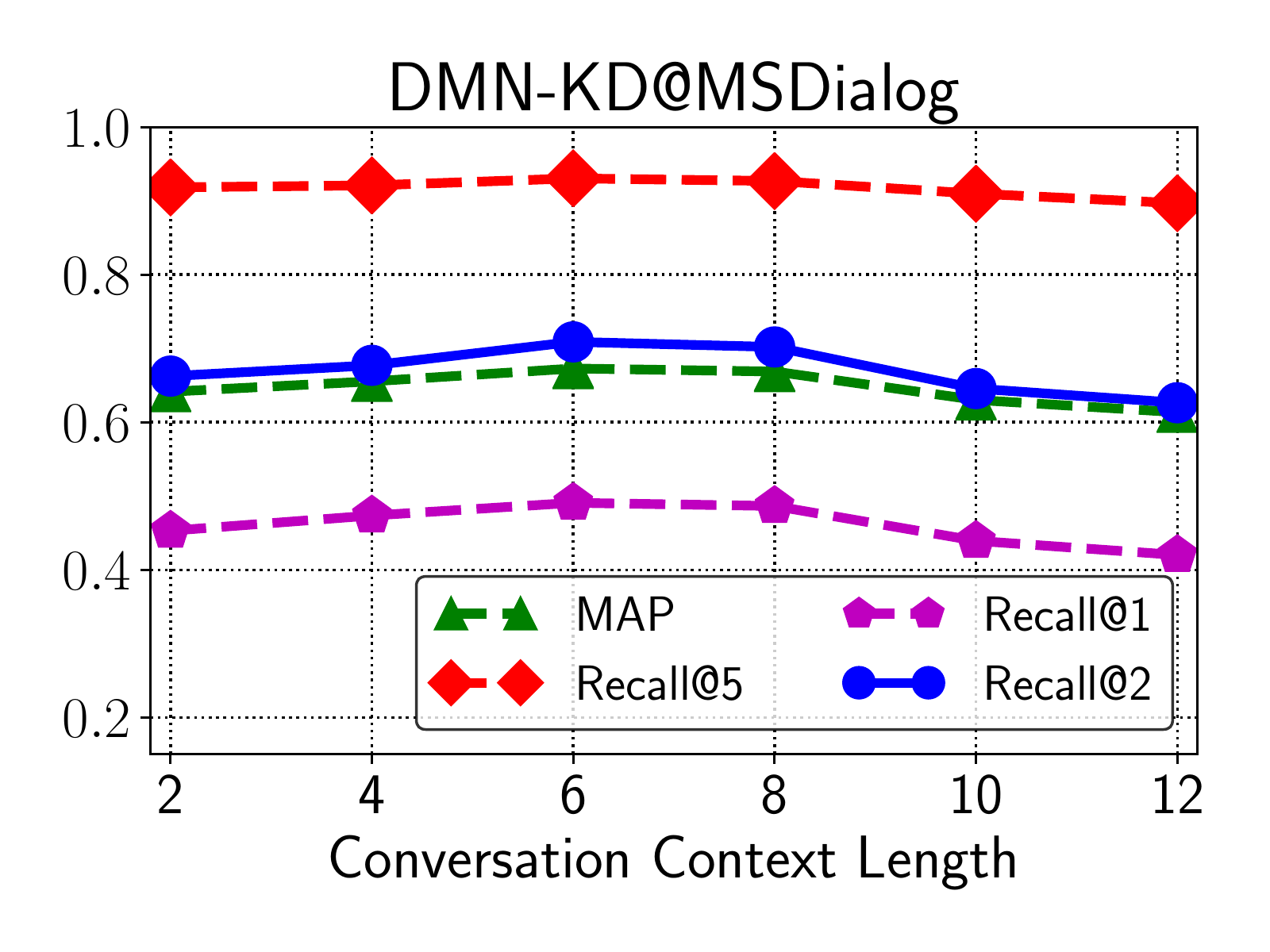}
		\label{fig:convlen-ms-dmnkd}
	\end{subfigure}
	\hspace{-0.1in}
	\begin{subfigure}[b]{0.245\textwidth}
		\includegraphics[width=\textwidth]{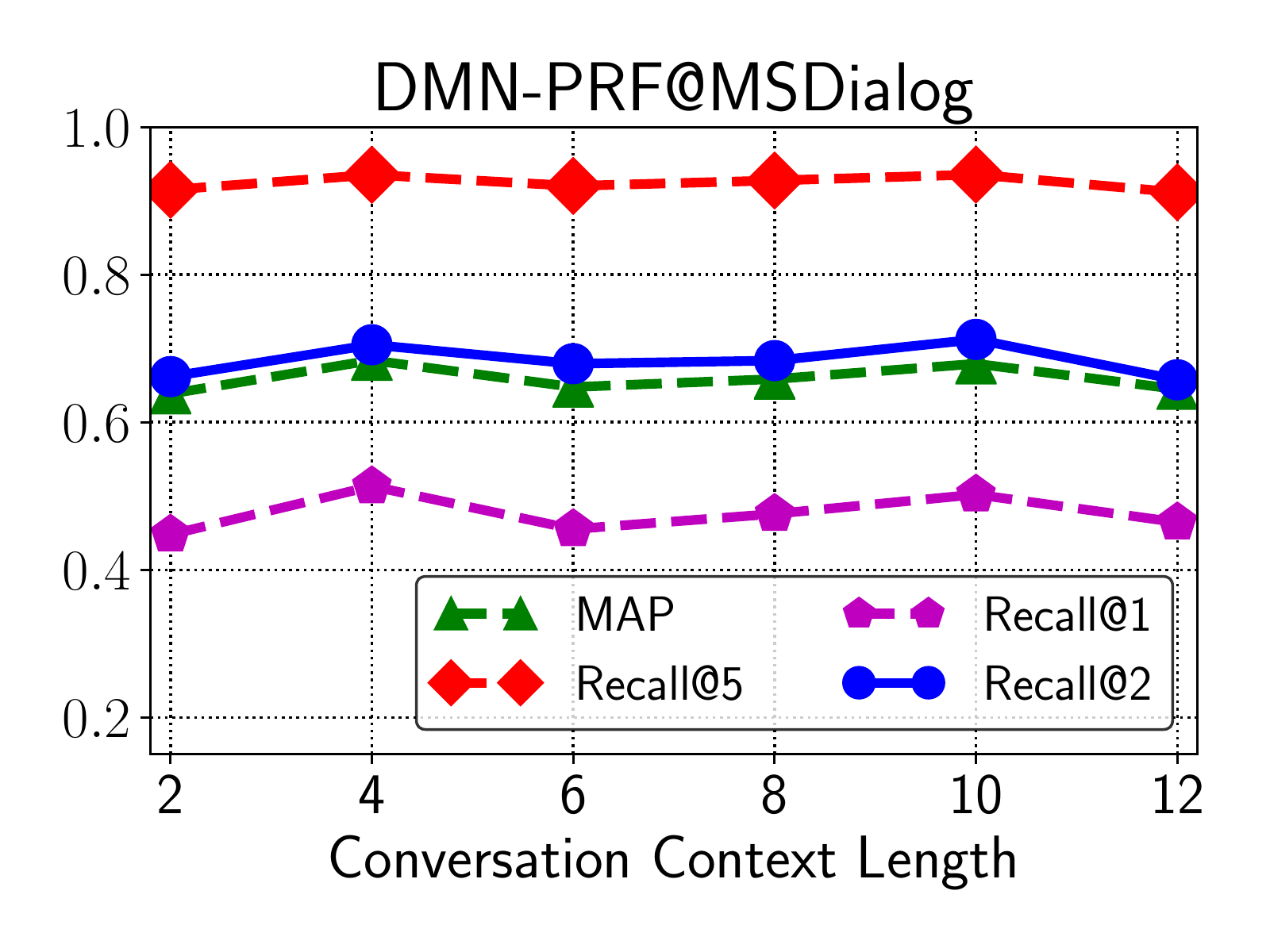}
		\label{fig:convlen-ms-dmnprf}
	\end{subfigure}
	\vspace{-0.25in}
	\caption{Performance of DMN-KD and DMN-PRF with different choices of conversation context length.}\label{fig:tune_convlen}
\end{figure*}

\subsubsection{\textbf{Performance Comparison over Different Response Types}}
%\ylcomment{to be updated. Update to the comparison of DMN-PRF/DMN-KD over DMN.}
We conduct fine-grained analysis on the performance of different models on different response types. We annotated the user intents in $10,020$ MSDialog utterances using Amazon Mechanical Turk \footnote{\url{https://www.mturk.com/}}. We defined $12$ user intent types including several types related to ``questions'' (original question, follow-up question, information request, clarifying question, and etc.), ``answers'' ( potential answer and further details), ``gratitude'' (expressing thanks, greetings) and ``feedback'' (positive feedback and negative feedback). Then we trained a Random Forest classifier with TF-IDF features and applied this classifier to predict the response candidate types in the testing data of MSDialog.  The dialog contexts were grouped by the type of the true response candidate. Finally we computed the average Recall@1 over different groups. Figure \ref{fig:result_diff_response_type} shows the results. We find that both DMN-KD and DMN-PRF improve the performances of SMN for responses with type ``questions'', ``answers'' and ``gratitude''. This indicates that incorporating external knowledge with PRF or QA correspondence knowledge distillation can help the model select better responses, especially for QA related responses. For responses with type ``Feedback'', DMN-KD and DMN-PRF achieved similar performances comparing with SMN.

\subsection{Model Ablation Analysis}
We investigate the effectiveness of different components of DMN-PRF and DMN-KD by removing them one by one from the original model with UDC and MSDialog data. We also study the effectiveness of different interaction types for $\mathbf{M1}/\mathbf{M2}/\mathbf{M3}$. Table \ref{tab:model_ablation_results} shows the results. We summarize our observations as follows: 1) For the interaction matrices, we find that the performance will drop if we remove any one of $\mathbf{M1}/\mathbf{M2}$ for DMN-PRF or $\mathbf{M1}/\mathbf{M2}/\mathbf{M3}$ for DMN-KD. This indicates that all of word level interaction matching, sequence level interaction matching and external QA correspondence interaction matching are useful for response selection in information-seeking conversation. 2) For interaction types, we can find that dot product is the best setting on both UDC and MSDialog except the results of DMN-KD on MSDialog. The next best one is cosine similarity. Bilinear product is the worst, especially on MSDialog data. This is because bilinear product will introduce a transformation matrix $\mathbf{A}$ as an additional model parameter, leading to higher model complexity. Thus the model is more likely to overfit the training data, especially for the relatively small MSDialog data. 3) If we only leave one channel in the interaction matrices, we can find that $\mathbf{M1}$ is more powerful than $\mathbf{M2}$ for DMN-PRF. For DMN-KD, $\mathbf{M1}$ is also the best one, followed by $\mathbf{M2}$. $\mathbf{M3}$ is the last one, but it stills adds additional matching signals when it is combined with $\mathbf{M1}$ and $\mathbf{M2}$. The matching signals $\mathbf{M3}$ from external collection could be supplementary features to the word embedding based matching matrix $\mathbf{M1}$ and BiGRU representation based matching matrix  $\mathbf{M2}$.

\subsection{\textbf{Impact of Conversation Context Length}}
%\ylcomment{Experiments with changing ``text1\_max\_utt\_num'' in the model configuration file. Draw figures on the change of performances $\{MAP, Recall@1, Recall@2, Recall@5\}$ of DMN-PRF and DMN-KD as the conversation context length change in $\{2,4,6,8,10,12\}$ for MSDialog and $\{3,6,9,12,15, 18\}$ for UDC.}
%\ylcomment{4 Figures}
We further analyze the impact of the conversation context length on the performances of our proposed DMN-KD and DMN-PRF models. 
%We conduct experiments by observing the change of model performances when we vary the choices of the maximum conversation context length in the model. We present the results in Figure \ref{fig:tune_convlen}. 
As presented in Figure \ref{fig:tune_convlen}, we find the performance first increases and then decreases, with the increase of conversation context length. The reason for these trends is that the context length controls the available previous utterances in the dialog context modeled by DMN-KD and DMN-PRF. If the context length is too small, there would be not enough information for the model to learn the matching patterns between the context and response candidates. However, setting the context length too large will also bring noise into the model results, since the words in utterances a few turns ago could be very different due to the topic changes during conversations. %\ylcomment{Add `` $8$ is a good choice for both data.''}

\subsection{Case Study}

% \usepackage{multirow}
\begin{table}[]
	\footnotesize
	\centering
	\caption{Examples of Top-1 ranked responses by different methods. $y_i^k$  means the label of a response candidate.}
	\vspace{-0.1in}
	\label{tab:case_study}
	\begin{tabular}{p{1.1cm}  | p{0.2cm} | p{6.3cm}}
		\hline  \hline% What is your computer set to for it's global date formatting?
		\multirow{2}{*}{Context} & \multicolumn{2}{p{6.9cm}}{ [User] I open Excel and it automatically formats my dates into American formatting. I have changed and saved the formatting to NZ style. } \\ %\cline{2-4} 
		& \multicolumn{2}{p{6.9cm}}{However everytime I pull the document out of office 365 it reverts back to the American format. How do I stop this ? \quad [Agent] Is it one file  or all files in Excel ? \quad [User] It does seem to be all Excel files. How do I change the global  date format setting ? }     \\  \hline
		Method      & $y_i^k$   & Top-1 Ranked Response        \\ \hline
		SMN     & 0    & Go to Settings -\textgreater System -\textgreater Tablet Mode....Change setting as indicated in the snapshot below.         \\ \hline
		DMN-KD   & 1      & That is a Windows setting.  Go to Control Panel \textgreater Regional settings.  This will change date settings for all applications.   \\ \hline
		DMN-PRF    & 1   & That is a Windows setting.  Go to Control Panel \textgreater Regional settings.  This will change date settings for all applications.\\ \hline \hline
	\end{tabular}
\end{table}

We perform a case study in Table \ref{tab:case_study} on the top ranked responses by different methods including SMN, DMN-KD and DMN-PRF. In this example, both DMN-KD and DMN-PRF produced correct top ranked responses. We checked the retrieved QA posts by the correct response candidate and found that ``\textit{settings, regional, change, windows, separator, format, excel, panel, application}'' are the most frequent terms. Among them ``\textit{excel}'' is especially useful for promoting the rank of the correct response candidate, since this term which is included multiple times by the dialog context does not actually appear in the raw text of the correct response candidate. This gives an example of the effectiveness of incorporating external knowledge from the retrieved QA posts into response candidates.

%\ylcomment{Doing this with one of MSDialog/UDC is enough.}
%
%\ylcomment{Do this with significant test together since both tasks need model prediction and model score files on testing data.}
%
%\ylcomment{Select some ranking examples to show on which cases DMN-PRF/DMN-KD/SMN is better/worse. Show example query dialog context where DMN-PRF/DMN-KD are better than DMN/SMN and explain why. Explain why the incorporated external knowledge in DMN-PRF/DMN-KD could be helpful in this example. Fine-grained/in-depth analysis to provide more insights and interesting findings. Refer to discussion notes with JF and examples in ECIR16 papers/NEUIR17 papers/ CX papers/ Cortana information cards Before\&After examples.}
%
%\ylcomment{Win/Tie/Loss analysis between DMN-PRF/DMN-KD/DMN over SMN. Refer to Table 6 and 7 of this paper \url{http://www.cs.cmu.edu/~cx/papers/word-entity-duet.pdf}}
%
%\ylcomment{Refer to Table 5 of this paper \url{https://arxiv.org/pdf/1506.06714.pdf}}
%
%\ylcomment{1 Table} \ylcomment{Adding more examples is good.}

%\ylcomment{Remove this section if there is no time}
%Hyper-parameter tuning results.
%
%More detailed experimental results.

% The paper should not exceed ten (10) pages in the current ACM two-column conference format (including references and figures). 

%\section{Discussion}
\section{Conclusions and Future Work}
\label{sec:conclu}

In this paper, we propose a learning framework based on deep matching networks to leverage external knowledge for response ranking in information-seeking conversation systems. We incorporate external knowledge into deep neural models with pseudo-relevance feedback and QA correspondence knowledge distillation. Extensive experiments %with information-seeking conversation data sets including both 
on both open benchmarks and commercial data show our methods outperform various baselines including the state-of-the-art methods. %on response selection in multi-turn conversations. 
We also perform analysis on different response types and model variations to provide insights on model applications. For future work, we plan to model user intent in information-seeking conversations and learn meaningful patterns from user intent dynamics to help response selection. Incorporating both structured and unstructured knowledge into deep matching networks for response ranking is also interesting to explore. %an interesting direction to explore. % in conversations.

%\ylcomment{Try to finish a complete version soon before 22:00 pm,  01/24 and send the draft to JF/Bruce for comments and revising.}

\section{Acknowledgments}
This work was supported in part by the Center for Intelligent Information Retrieval and in part by NSF grant \#IIS-1419693. Any opinions, findings and conclusions or recommendations expressed in this material are those of the authors and do not necessarily reflect those of the sponsor.

\bibliographystyle{ACM-Reference-Format}
\bibliography{reference}


\begin{thebibliography}{48}

%%% the Web address.

\ifx \showCODEN    \undefined \def \showCODEN     #1{\unskip}     \fi
\ifx \showDOI      \undefined \def \showDOI       #1{#1}\fi
\ifx \showISBNx    \undefined \def \showISBNx     #1{\unskip}     \fi
\ifx \showISBNxiii \undefined \def \showISBNxiii  #1{\unskip}     \fi
\ifx \showISSN     \undefined \def \showISSN      #1{\unskip}     \fi
\ifx \showLCCN     \undefined \def \showLCCN      #1{\unskip}     \fi
\ifx \shownote     \undefined \def \shownote      #1{#1}          \fi
\ifx \showarticletitle \undefined \def \showarticletitle #1{#1}   \fi
\ifx \showURL      \undefined \def \showURL       {\relax}        \fi
% invisible to TeX
\providecommand\bibfield[2]{#2}
\providecommand\bibinfo[2]{#2}
\providecommand\natexlab[1]{#1}
\providecommand\showeprint[2][]{arXiv:#2}

\bibitem[\protect\citeauthoryear{Arguello, Choi, and Capra}{Arguello
  et~al\mbox{.}}{2017}]%
        {arguello2017factors}
\bibfield{author}{\bibinfo{person}{J. Arguello}, \bibinfo{person}{B. Choi},
  {and} \bibinfo{person}{R. Capra}.} \bibinfo{year}{2017}\natexlab{}.
\newblock \showarticletitle{Factors Affecting Users' Information Requests}. In
  \bibinfo{booktitle}{\emph{CAIR '17}}.
\newblock


\bibitem[\protect\citeauthoryear{Bordes, Boureau, and Weston}{Bordes
  et~al\mbox{.}}{2017}]%
        {bordes2017learning}
\bibfield{author}{\bibinfo{person}{A. Bordes}, \bibinfo{person}{Y. Boureau},
  {and} \bibinfo{person}{J. Weston}.} \bibinfo{year}{2017}\natexlab{}.
\newblock \showarticletitle{Learning end-to-end goal-oriented dialog}.
\newblock \bibinfo{journal}{\emph{ICLR '17}}.
\newblock


\bibitem[\protect\citeauthoryear{Cao, Nie, Gao, and Stephen}{Cao
  et~al\mbox{.}}{2008}]%
        {Cao:2008:SGE:1390334.1390377}
\bibfield{author}{\bibinfo{person}{G. Cao}, \bibinfo{person}{J. Nie},
  \bibinfo{person}{J. Gao}, {and} \bibinfo{person}{R. Stephen}.}
  \bibinfo{year}{2008}\natexlab{}.
\newblock \showarticletitle{Selecting Good Expansion Terms for Pseudo-relevance
  Feedback}. In \bibinfo{booktitle}{\emph{SIGIR '08}}.
\newblock


\bibitem[\protect\citeauthoryear{Chung, G{\"{u}}l{\c{c}}ehre, Cho, and
  Bengio}{Chung et~al\mbox{.}}{2014}]%
        {DBLP:journals/corr/ChungGCB14}
\bibfield{author}{\bibinfo{person}{J. Chung}, \bibinfo{person}{{\c{C}}.
  G{\"{u}}l{\c{c}}ehre}, \bibinfo{person}{K. Cho}, {and} \bibinfo{person}{Y.
  Bengio}.} \bibinfo{year}{2014}\natexlab{}.
\newblock \showarticletitle{Empirical Evaluation of Gated Recurrent Neural
  Networks on Sequence Modeling}.
\newblock \bibinfo{journal}{\emph{CoRR}} (\bibinfo{year}{2014}).
\newblock


\bibitem[\protect\citeauthoryear{Dhingra, Li, Li, Gao, Chen, Ahmed, and
  Deng}{Dhingra et~al\mbox{.}}{2017}]%
        {P17-1045}
\bibfield{author}{\bibinfo{person}{B. Dhingra}, \bibinfo{person}{L. Li},
  \bibinfo{person}{X. Li}, \bibinfo{person}{J. Gao}, \bibinfo{person}{Y. Chen},
  \bibinfo{person}{F. Ahmed}, {and} \bibinfo{person}{L. Deng}.}
  \bibinfo{year}{2017}\natexlab{}.
\newblock \showarticletitle{Towards End-to-End Reinforcement Learning of
  Dialogue Agents for Information Access}. In
  \bibinfo{booktitle}{\emph{ACL'17}}.
\newblock


\bibitem[\protect\citeauthoryear{Guo, Fan, Ai, and Croft}{Guo
  et~al\mbox{.}}{2016}]%
        {Guo:2016:DRM:2983323.2983769}
\bibfield{author}{\bibinfo{person}{J. Guo}, \bibinfo{person}{Y. Fan},
  \bibinfo{person}{Q. Ai}, {and} \bibinfo{person}{W.~B. Croft}.}
  \bibinfo{year}{2016}\natexlab{}.
\newblock \showarticletitle{A Deep Relevance Matching Model for Ad-hoc
  Retrieval}. In \bibinfo{booktitle}{\emph{CIKM '16}}.
\newblock


\bibitem[\protect\citeauthoryear{Hu, Lu, Li, and Chen}{Hu
  et~al\mbox{.}}{2014}]%
        {DBLP:conf/nips/HuLLC14}
\bibfield{author}{\bibinfo{person}{B. Hu}, \bibinfo{person}{Z. Lu},
  \bibinfo{person}{H. Li}, {and} \bibinfo{person}{Q. Chen}.}
  \bibinfo{year}{2014}\natexlab{}.
\newblock \showarticletitle{Convolutional Neural Network Architectures for
  Matching Natural Language Sentences}. In \bibinfo{booktitle}{\emph{NIPS
  '14}}.
\newblock


\bibitem[\protect\citeauthoryear{Huang, He, Gao, Deng, Acero, and Heck}{Huang
  et~al\mbox{.}}{2013}]%
        {DBLP:conf/cikm/HuangHGDAH13}
\bibfield{author}{\bibinfo{person}{P. Huang}, \bibinfo{person}{X. He},
  \bibinfo{person}{J. Gao}, \bibinfo{person}{L. Deng}, \bibinfo{person}{A.
  Acero}, {and} \bibinfo{person}{L.~P. Heck}.} \bibinfo{year}{2013}\natexlab{}.
\newblock \showarticletitle{Learning deep structured semantic models for web
  search using clickthrough data}. In \bibinfo{booktitle}{\emph{CIKM '13}}.
\newblock


\bibitem[\protect\citeauthoryear{Ji, Lu, and Li}{Ji et~al\mbox{.}}{2014}]%
        {DBLP:journals/corr/JiLL14}
\bibfield{author}{\bibinfo{person}{Z. Ji}, \bibinfo{person}{Z. Lu}, {and}
  \bibinfo{person}{H. Li}.} \bibinfo{year}{2014}\natexlab{}.
\newblock \showarticletitle{An Information Retrieval Approach to Short Text
  Conversation}.
\newblock \bibinfo{journal}{\emph{CoRR}}  \bibinfo{volume}{abs/1408.6988}
  (\bibinfo{year}{2014}).
\newblock


\bibitem[\protect\citeauthoryear{Kenter and de~Rijke}{Kenter and
  de~Rijke}{2017}]%
        {kenter-attentive-2017}
\bibfield{author}{\bibinfo{person}{T. Kenter} {and} \bibinfo{person}{M. de
  Rijke}.} \bibinfo{year}{2017}\natexlab{}.
\newblock \showarticletitle{Attentive memory networks: Efficient machine
  reading for conversational search}. In \bibinfo{booktitle}{\emph{CAIR '17}}.
\newblock


\bibitem[\protect\citeauthoryear{Kingma and Ba}{Kingma and Ba}{2014}]%
        {DBLP:journals/corr/KingmaB14}
\bibfield{author}{\bibinfo{person}{D.~P. Kingma} {and} \bibinfo{person}{J.
  Ba}.} \bibinfo{year}{2014}\natexlab{}.
\newblock \showarticletitle{Adam: {A} Method for Stochastic Optimization}.
\newblock \bibinfo{journal}{\emph{CoRR}} (\bibinfo{year}{2014}).
\newblock


\bibitem[\protect\citeauthoryear{Lavrenko and Croft}{Lavrenko and
  Croft}{2001}]%
        {Lavrenko:2001:RBL:383952.383972}
\bibfield{author}{\bibinfo{person}{V. Lavrenko} {and} \bibinfo{person}{W.~B.
  Croft}.} \bibinfo{year}{2001}\natexlab{}.
\newblock \showarticletitle{Relevance Based Language Models}. In
  \bibinfo{booktitle}{\emph{SIGIR '01}}.
\newblock


\bibitem[\protect\citeauthoryear{Li, Qiu, Chen, Wang, Gao, Huang, Ren, Zhao,
  Zhao, Wang, and Jin}{Li et~al\mbox{.}}{2017}]%
        {alime-demo}
\bibfield{author}{\bibinfo{person}{F. Li}, \bibinfo{person}{M. Qiu},
  \bibinfo{person}{H. Chen}, \bibinfo{person}{X. Wang}, \bibinfo{person}{X.
  Gao}, \bibinfo{person}{J. Huang}, \bibinfo{person}{J. Ren},
  \bibinfo{person}{Z. Zhao}, \bibinfo{person}{W. Zhao}, \bibinfo{person}{L.
  Wang}, {and} \bibinfo{person}{G. Jin}.} \bibinfo{year}{2017}\natexlab{}.
\newblock \showarticletitle{AliMe Assist: An Intelligent Assistant for Creating
  an Innovative E-commerce Experience}. In \bibinfo{booktitle}{\emph{CIKM
  '17}}.
\newblock


\bibitem[\protect\citeauthoryear{Li, Galley, Brockett, Spithourakis, Gao, and
  Dolan}{Li et~al\mbox{.}}{2016a}]%
        {DBLP:conf/acl/LiGBSGD16}
\bibfield{author}{\bibinfo{person}{J. Li}, \bibinfo{person}{M. Galley},
  \bibinfo{person}{C. Brockett}, \bibinfo{person}{G.~P. Spithourakis},
  \bibinfo{person}{J. Gao}, {and} \bibinfo{person}{W.~B. Dolan}.}
  \bibinfo{year}{2016}\natexlab{a}.
\newblock \showarticletitle{A Persona-Based Neural Conversation Model}. In
  \bibinfo{booktitle}{\emph{ACL'16}}.
\newblock


\bibitem[\protect\citeauthoryear{Li, Monroe, Ritter, Jurafsky, Galley, and
  Gao}{Li et~al\mbox{.}}{2016b}]%
        {DBLP:conf/emnlp/LiMRJGG16}
\bibfield{author}{\bibinfo{person}{J. Li}, \bibinfo{person}{W. Monroe},
  \bibinfo{person}{A. Ritter}, \bibinfo{person}{D. Jurafsky},
  \bibinfo{person}{M. Galley}, {and} \bibinfo{person}{J. Gao}.}
  \bibinfo{year}{2016}\natexlab{b}.
\newblock \showarticletitle{Deep Reinforcement Learning for Dialogue
  Generation}. In \bibinfo{booktitle}{\emph{EMNLP'16}}.
\newblock


\bibitem[\protect\citeauthoryear{Lowe, Pow, Serban, and Pineau}{Lowe
  et~al\mbox{.}}{2015}]%
        {DBLP:journals/corr/LowePSP15}
\bibfield{author}{\bibinfo{person}{R. Lowe}, \bibinfo{person}{N. Pow},
  \bibinfo{person}{I. Serban}, {and} \bibinfo{person}{J. Pineau}.}
  \bibinfo{year}{2015}\natexlab{}.
\newblock \showarticletitle{The Ubuntu Dialogue Corpus: {A} Large Dataset for
  Research in Unstructured Multi-Turn Dialogue Systems}.
\newblock \bibinfo{journal}{\emph{CoRR}}  \bibinfo{volume}{abs/1506.08909}
  (\bibinfo{year}{2015}).
\newblock


\bibitem[\protect\citeauthoryear{Lv and Zhai}{Lv and Zhai}{2009}]%
        {Lv:2009:CSM:1645953.1646259}
\bibfield{author}{\bibinfo{person}{Y. Lv} {and} \bibinfo{person}{C. Zhai}.}
  \bibinfo{year}{2009}\natexlab{}.
\newblock \showarticletitle{A Comparative Study of Methods for Estimating Query
  Language Models with Pseudo Feedback}. In \bibinfo{booktitle}{\emph{CIKM
  '09}}.
\newblock


\bibitem[\protect\citeauthoryear{Mikolov, Sutskever, Chen, Corrado, and
  Dean}{Mikolov et~al\mbox{.}}{2013}]%
        {DBLP:conf/nips/MikolovSCCD13}
\bibfield{author}{\bibinfo{person}{T. Mikolov}, \bibinfo{person}{I. Sutskever},
  \bibinfo{person}{K. Chen}, \bibinfo{person}{G.~S. Corrado}, {and}
  \bibinfo{person}{J. Dean}.} \bibinfo{year}{2013}\natexlab{}.
\newblock \showarticletitle{Distributed Representations of Words and Phrases
  and their Compositionality}. In \bibinfo{booktitle}{\emph{NIPS '13}}.
\newblock


\bibitem[\protect\citeauthoryear{Mitra, Diaz, and Craswell}{Mitra
  et~al\mbox{.}}{2017}]%
        {Mitra:2017:LMU:3038912.3052579}
\bibfield{author}{\bibinfo{person}{B. Mitra}, \bibinfo{person}{F. Diaz}, {and}
  \bibinfo{person}{N. Craswell}.} \bibinfo{year}{2017}\natexlab{}.
\newblock \showarticletitle{Learning to Match Using Local and Distributed
  Representations of Text for Web Search}. In \bibinfo{booktitle}{\emph{WWW
  '17}}.
\newblock


\bibitem[\protect\citeauthoryear{Pang, Lan, Guo, Xu, Wan, and Cheng}{Pang
  et~al\mbox{.}}{2016}]%
        {DBLP:conf/aaai/PangLGXWC16}
\bibfield{author}{\bibinfo{person}{L. Pang}, \bibinfo{person}{Y. Lan},
  \bibinfo{person}{J. Guo}, \bibinfo{person}{J. Xu}, \bibinfo{person}{S. Wan},
  {and} \bibinfo{person}{X. Cheng}.} \bibinfo{year}{2016}\natexlab{}.
\newblock \showarticletitle{Text Matching as Image Recognition}. In
  \bibinfo{booktitle}{\emph{AAAI '16}}.
\newblock


\bibitem[\protect\citeauthoryear{Qiu, Li, Wang, Gao, Chen, Zhao, Chen, Huang,
  and Chu}{Qiu et~al\mbox{.}}{2017}]%
        {alime-chat}
\bibfield{author}{\bibinfo{person}{M. Qiu}, \bibinfo{person}{F. Li},
  \bibinfo{person}{S. Wang}, \bibinfo{person}{X. Gao}, \bibinfo{person}{Y.
  Chen}, \bibinfo{person}{W. Zhao}, \bibinfo{person}{H. Chen},
  \bibinfo{person}{J. Huang}, {and} \bibinfo{person}{W. Chu}.}
  \bibinfo{year}{2017}\natexlab{}.
\newblock \showarticletitle{AliMe Chat: A Sequence to Sequence and Rerank based
  Chatbot Engine}. In \bibinfo{booktitle}{\emph{ACL '17}}.
\newblock


\bibitem[\protect\citeauthoryear{Radlinski and Craswell}{Radlinski and
  Craswell}{2017}]%
        {radlinski2017theoretical}
\bibfield{author}{\bibinfo{person}{F. Radlinski} {and} \bibinfo{person}{N.
  Craswell}.} \bibinfo{year}{2017}\natexlab{}.
\newblock \showarticletitle{A theoretical framework for conversational search}.
  In \bibinfo{booktitle}{\emph{CHIIR '17}}.
\newblock


\bibitem[\protect\citeauthoryear{Ritter, Cherry, and Dolan}{Ritter
  et~al\mbox{.}}{2011}]%
        {DBLP:conf/emnlp/RitterCD11}
\bibfield{author}{\bibinfo{person}{A. Ritter}, \bibinfo{person}{C. Cherry},
  {and} \bibinfo{person}{W.~B. Dolan}.} \bibinfo{year}{2011}\natexlab{}.
\newblock \showarticletitle{Data-Driven Response Generation in Social Media}.
  In \bibinfo{booktitle}{\emph{ACL '11}}.
\newblock


\bibitem[\protect\citeauthoryear{Robertson and Walker}{Robertson and
  Walker}{1994}]%
        {Robertson:1994:SEA:188490.188561}
\bibfield{author}{\bibinfo{person}{S. Robertson} {and} \bibinfo{person}{S.
  Walker}.} \bibinfo{year}{1994}\natexlab{}.
\newblock \showarticletitle{Some Simple Effective Approximations to the
  2-Poisson Model for Probabilistic Weighted Retrieval}. In
  \bibinfo{booktitle}{\emph{SIGIR '94}}.
\newblock


\bibitem[\protect\citeauthoryear{Rocchio}{Rocchio}{1971}]%
        {rocchio71relevance}
\bibfield{author}{\bibinfo{person}{J.~J. Rocchio}.}
  \bibinfo{year}{1971}\natexlab{}.
\newblock \showarticletitle{Relevance feedback in information retrieval}.
\newblock In \bibinfo{booktitle}{\emph{The Smart retrieval system - experiments
  in automatic document processing}},
  \bibfield{editor}{\bibinfo{person}{G.~Salton}} (Ed.).
\newblock


\bibitem[\protect\citeauthoryear{Shang, Lu, and Li}{Shang
  et~al\mbox{.}}{2015}]%
        {DBLP:conf/acl/ShangLL15}
\bibfield{author}{\bibinfo{person}{L. Shang}, \bibinfo{person}{Z. Lu}, {and}
  \bibinfo{person}{H. Li}.} \bibinfo{year}{2015}\natexlab{}.
\newblock \showarticletitle{Neural Responding Machine for Short-Text
  Conversation}. In \bibinfo{booktitle}{\emph{ACL '15}}.
\newblock


\bibitem[\protect\citeauthoryear{Sordoni, Galley, Auli, Brockett, Ji, Mitchell,
  Nie, Gao, and Dolan}{Sordoni et~al\mbox{.}}{2015}]%
        {DBLP:conf/naacl/SordoniGABJMNGD15}
\bibfield{author}{\bibinfo{person}{A. Sordoni}, \bibinfo{person}{M. Galley},
  \bibinfo{person}{M. Auli}, \bibinfo{person}{C. Brockett}, \bibinfo{person}{Y.
  Ji}, \bibinfo{person}{M. Mitchell}, \bibinfo{person}{J. Nie},
  \bibinfo{person}{J. Gao}, {and} \bibinfo{person}{B. Dolan}.}
  \bibinfo{year}{2015}\natexlab{}.
\newblock \showarticletitle{A Neural Network Approach to Context-Sensitive
  Generation of Conversational Responses}. In \bibinfo{booktitle}{\emph{NAACL
  '15}}.
\newblock


\bibitem[\protect\citeauthoryear{Spina, Trippas, Cavedon, and Sanderson}{Spina
  et~al\mbox{.}}{2017}]%
        {spina2017extracting}
\bibfield{author}{\bibinfo{person}{D. Spina}, \bibinfo{person}{J.~R Trippas},
  \bibinfo{person}{L. Cavedon}, {and} \bibinfo{person}{M. Sanderson}.}
  \bibinfo{year}{2017}\natexlab{}.
\newblock \showarticletitle{Extracting audio summaries to support effective
  spoken document search}.
\newblock \bibinfo{journal}{\emph{JAIST '17}} \bibinfo{volume}{68},
  \bibinfo{number}{9} (\bibinfo{year}{2017}).
\newblock


\bibitem[\protect\citeauthoryear{Thomas, McDu, Czerwinski, and Craswell}{Thomas
  et~al\mbox{.}}{2017}]%
        {thomas2017misc}
\bibfield{author}{\bibinfo{person}{P. Thomas}, \bibinfo{person}{D. McDu},
  \bibinfo{person}{M. Czerwinski}, {and} \bibinfo{person}{N. Craswell}.}
  \bibinfo{year}{2017}\natexlab{}.
\newblock \showarticletitle{MISC: A data set of information-seeking
  conversations}. In \bibinfo{booktitle}{\emph{CAIR '17}}.
\newblock


\bibitem[\protect\citeauthoryear{Tian, Yan, Mou, Song, Feng, and Zhao}{Tian
  et~al\mbox{.}}{2017}]%
        {DBLP:conf/acl/TianYMSFZ17}
\bibfield{author}{\bibinfo{person}{Z. Tian}, \bibinfo{person}{R. Yan},
  \bibinfo{person}{L. Mou}, \bibinfo{person}{Y. Song}, \bibinfo{person}{Y.
  Feng}, {and} \bibinfo{person}{D. Zhao}.} \bibinfo{year}{2017}\natexlab{}.
\newblock \showarticletitle{How to Make Context More Useful? An Empirical Study
  on Context-Aware Neural Conversational Models}. In
  \bibinfo{booktitle}{\emph{ACL '17}}.
\newblock


\bibitem[\protect\citeauthoryear{Trippas, Spina, Sanderson, and
  Cavedon}{Trippas et~al\mbox{.}}{2015}]%
        {trippas2015towards}
\bibfield{author}{\bibinfo{person}{J. Trippas}, \bibinfo{person}{D. Spina},
  \bibinfo{person}{M. Sanderson}, {and} \bibinfo{person}{L. Cavedon}.}
  \bibinfo{year}{2015}\natexlab{}.
\newblock \showarticletitle{Towards understanding the impact of length in web
  search result summaries over a speech-only communication channel}. In
  \bibinfo{booktitle}{\emph{SIGIR '15}}.
\newblock


\bibitem[\protect\citeauthoryear{Vinyals and Le}{Vinyals and Le}{2015}]%
        {DBLP:journals/corr/VinyalsL15}
\bibfield{author}{\bibinfo{person}{O. Vinyals} {and} \bibinfo{person}{Q.~V.
  Le}.} \bibinfo{year}{2015}\natexlab{}.
\newblock \showarticletitle{A Neural Conversational Model}.
\newblock \bibinfo{journal}{\emph{CoRR}}  \bibinfo{volume}{abs/1506.05869}
  (\bibinfo{year}{2015}).
\newblock


\bibitem[\protect\citeauthoryear{Wan, Lan, Guo, Xu, Pang, and Cheng}{Wan
  et~al\mbox{.}}{2016}]%
        {DBLP:conf/aaai/WanLGXPC16}
\bibfield{author}{\bibinfo{person}{S. Wan}, \bibinfo{person}{Y. Lan},
  \bibinfo{person}{J. Guo}, \bibinfo{person}{J. Xu}, \bibinfo{person}{L. Pang},
  {and} \bibinfo{person}{X. Cheng}.} \bibinfo{year}{2016}\natexlab{}.
\newblock \showarticletitle{A Deep Architecture for Semantic Matching with
  Multiple Positional Sentence Representations}. In
  \bibinfo{booktitle}{\emph{AAAI '16}}.
\newblock


\bibitem[\protect\citeauthoryear{Wang, Lu, Li, and Chen}{Wang
  et~al\mbox{.}}{2013}]%
        {DBLP:conf/emnlp/WangLLC13}
\bibfield{author}{\bibinfo{person}{H. Wang}, \bibinfo{person}{Z. Lu},
  \bibinfo{person}{H. Li}, {and} \bibinfo{person}{E. Chen}.}
  \bibinfo{year}{2013}\natexlab{}.
\newblock \showarticletitle{A Dataset for Research on Short-Text
  Conversations}. In \bibinfo{booktitle}{\emph{EMNLP '13}}.
\newblock


\bibitem[\protect\citeauthoryear{Wen, Vandyke, Mrksic, Gasic, Rojas-Barahona,
  Su, Ultes, and Young}{Wen et~al\mbox{.}}{2017}]%
        {wen2016network}
\bibfield{author}{\bibinfo{person}{T. Wen}, \bibinfo{person}{D. Vandyke},
  \bibinfo{person}{N. Mrksic}, \bibinfo{person}{M. Gasic},
  \bibinfo{person}{L.~M Rojas-Barahona}, \bibinfo{person}{P. Su},
  \bibinfo{person}{S. Ultes}, {and} \bibinfo{person}{S. Young}.}
  \bibinfo{year}{2017}\natexlab{}.
\newblock \showarticletitle{A network-based end-to-end trainable task-oriented
  dialogue system}.
\newblock \bibinfo{journal}{\emph{EACL '17}} (\bibinfo{year}{2017}).
\newblock


\bibitem[\protect\citeauthoryear{Wu, Wu, Xing, Zhou, and Li}{Wu
  et~al\mbox{.}}{2017}]%
        {DBLP:conf/acl/WuWXZL17}
\bibfield{author}{\bibinfo{person}{Y. Wu}, \bibinfo{person}{W. Wu},
  \bibinfo{person}{C. Xing}, \bibinfo{person}{M. Zhou}, {and}
  \bibinfo{person}{Z. Li}.} \bibinfo{year}{2017}\natexlab{}.
\newblock \showarticletitle{Sequential Matching Network: {A} New Architecture
  for Multi-turn Response Selection in Retrieval-Based Chatbots}. In
  \bibinfo{booktitle}{\emph{ACL '17}}.
\newblock


\bibitem[\protect\citeauthoryear{Xiong, Dai, Callan, Liu, and Power}{Xiong
  et~al\mbox{.}}{2017}]%
        {Xiong:2017:ENA:3077136.3080809}
\bibfield{author}{\bibinfo{person}{C. Xiong}, \bibinfo{person}{Z. Dai},
  \bibinfo{person}{J. Callan}, \bibinfo{person}{Z. Liu}, {and}
  \bibinfo{person}{R. Power}.} \bibinfo{year}{2017}\natexlab{}.
\newblock \showarticletitle{End-to-End Neural Ad-hoc Ranking with Kernel
  Pooling}. In \bibinfo{booktitle}{\emph{SIGIR '17}}.
\newblock


\bibitem[\protect\citeauthoryear{Xu, Liu, Wang, Sun, and Wang}{Xu
  et~al\mbox{.}}{2016}]%
        {DBLP:journals/corr/XuLWSW16}
\bibfield{author}{\bibinfo{person}{Z. Xu}, \bibinfo{person}{B. Liu},
  \bibinfo{person}{B. Wang}, \bibinfo{person}{C. Sun}, {and}
  \bibinfo{person}{X. Wang}.} \bibinfo{year}{2016}\natexlab{}.
\newblock \showarticletitle{Incorporating Loose-Structured Knowledge into
  {LSTM} with Recall Gate for Conversation Modeling}.
\newblock \bibinfo{journal}{\emph{CoRR}} (\bibinfo{year}{2016}).
\newblock


\bibitem[\protect\citeauthoryear{Yan, Song, and Wu}{Yan et~al\mbox{.}}{2016a}]%
        {DBLP:conf/sigir/YanSW16}
\bibfield{author}{\bibinfo{person}{R. Yan}, \bibinfo{person}{Y. Song}, {and}
  \bibinfo{person}{H. Wu}.} \bibinfo{year}{2016}\natexlab{a}.
\newblock \showarticletitle{Learning to Respond with Deep Neural Networks for
  Retrieval-Based Human-Computer Conversation System}. In
  \bibinfo{booktitle}{\emph{SIGIR}}.
\newblock


\bibitem[\protect\citeauthoryear{Yan, Song, Zhou, and Wu}{Yan
  et~al\mbox{.}}{2016b}]%
        {DBLP:conf/cikm/YanSZW16}
\bibfield{author}{\bibinfo{person}{R. Yan}, \bibinfo{person}{Y. Song},
  \bibinfo{person}{X. Zhou}, {and} \bibinfo{person}{H. Wu}.}
  \bibinfo{year}{2016}\natexlab{b}.
\newblock \showarticletitle{"Shall {I} Be Your Chat Companion?": Towards an
  Online Human-Computer Conversation System}. In \bibinfo{booktitle}{\emph{CIKM
  '16}}.
\newblock


\bibitem[\protect\citeauthoryear{Yan, Zhao, and E.}{Yan et~al\mbox{.}}{2017}]%
        {DBLP:conf/sigir/YanZE17}
\bibfield{author}{\bibinfo{person}{R. Yan}, \bibinfo{person}{D. Zhao}, {and}
  \bibinfo{person}{W. E.}} \bibinfo{year}{2017}\natexlab{}.
\newblock \showarticletitle{Joint Learning of Response Ranking and Next
  Utterance Suggestion in Human-Computer Conversation System}. In
  \bibinfo{booktitle}{\emph{SIGIR '17}}.
\newblock


\bibitem[\protect\citeauthoryear{Yang, Ai, Guo, and Croft}{Yang
  et~al\mbox{.}}{2016}]%
        {Yang:2016:ARS:2983323.2983818}
\bibfield{author}{\bibinfo{person}{L. Yang}, \bibinfo{person}{Q. Ai},
  \bibinfo{person}{J. Guo}, {and} \bibinfo{person}{W.~B. Croft}.}
  \bibinfo{year}{2016}\natexlab{}.
\newblock \showarticletitle{aNMM: Ranking Short Answer Texts with
  Attention-Based Neural Matching Model}. In \bibinfo{booktitle}{\emph{CIKM
  '16}}.
\newblock


\bibitem[\protect\citeauthoryear{Yang, Zamani, Zhang, Guo, and Croft}{Yang
  et~al\mbox{.}}{2017}]%
        {DBLP:journals/corr/YangZZGC17}
\bibfield{author}{\bibinfo{person}{L. Yang}, \bibinfo{person}{H. Zamani},
  \bibinfo{person}{Y. Zhang}, \bibinfo{person}{J. Guo}, {and}
  \bibinfo{person}{W.~B. Croft}.} \bibinfo{year}{2017}\natexlab{}.
\newblock \showarticletitle{Neural Matching Models for Question Retrieval and
  Next Question Prediction in Conversation}.
\newblock \bibinfo{journal}{\emph{CoRR}} (\bibinfo{year}{2017}).
\newblock


\bibitem[\protect\citeauthoryear{Young, Ga\v{s}i\'{c}, Keizer, Mairesse,
  Schatzmann, Thomson, and Yu}{Young et~al\mbox{.}}{2010}]%
        {Young:2010:HIS:1621140.1621240}
\bibfield{author}{\bibinfo{person}{S. Young}, \bibinfo{person}{M.
  Ga\v{s}i\'{c}}, \bibinfo{person}{S. Keizer}, \bibinfo{person}{F. Mairesse},
  \bibinfo{person}{J. Schatzmann}, \bibinfo{person}{B. Thomson}, {and}
  \bibinfo{person}{K. Yu}.} \bibinfo{year}{2010}\natexlab{}.
\newblock \showarticletitle{The Hidden Information State Model: A Practical
  Framework for POMDP-based Spoken Dialogue Management}.
\newblock \bibinfo{journal}{\emph{Comput. Speech Lang.}}
  (\bibinfo{year}{2010}).
\newblock


\bibitem[\protect\citeauthoryear{Yu, Qiu, Jiang, Huang, Song, Chu, and Chen}{Yu
  et~al\mbox{.}}{2018}]%
        {alime-tl}
\bibfield{author}{\bibinfo{person}{J. Yu}, \bibinfo{person}{M. Qiu},
  \bibinfo{person}{J. Jiang}, \bibinfo{person}{J. Huang}, \bibinfo{person}{S.
  Song}, \bibinfo{person}{W. Chu}, {and} \bibinfo{person}{H. Chen}.}
  \bibinfo{year}{2018}\natexlab{}.
\newblock \showarticletitle{Modelling Domain Relationships for Transfer
  Learning on Retrieval-based Question Answering Systems in E-commerce}.
\newblock \bibinfo{journal}{\emph{WSDM '18}}.
\newblock


\bibitem[\protect\citeauthoryear{Zamani, Dadashkarimi, Shakery, and
  Croft}{Zamani et~al\mbox{.}}{2016}]%
        {Zamani:2016:PFB:2983323.2983844}
\bibfield{author}{\bibinfo{person}{H. Zamani}, \bibinfo{person}{J.
  Dadashkarimi}, \bibinfo{person}{A. Shakery}, {and} \bibinfo{person}{W.~B.
  Croft}.} \bibinfo{year}{2016}\natexlab{}.
\newblock \showarticletitle{Pseudo-Relevance Feedback Based on Matrix
  Factorization}. In \bibinfo{booktitle}{\emph{CIKM '16}}.
\newblock


\bibitem[\protect\citeauthoryear{Zhai and Lafferty}{Zhai and Lafferty}{2001}]%
        {Zhai:2001:MFL:502585.502654}
\bibfield{author}{\bibinfo{person}{C. Zhai} {and} \bibinfo{person}{J.
  Lafferty}.} \bibinfo{year}{2001}\natexlab{}.
\newblock \showarticletitle{Model-based Feedback in the Language Modeling
  Approach to Information Retrieval}. In \bibinfo{booktitle}{\emph{CIKM '01}}.
\newblock


\bibitem[\protect\citeauthoryear{Zhou, Dong, Wu, Zhao, Yu, Tian, Liu, and
  Yan}{Zhou et~al\mbox{.}}{2016}]%
        {DBLP:conf/emnlp/ZhouDWZYTLY16}
\bibfield{author}{\bibinfo{person}{X. Zhou}, \bibinfo{person}{D. Dong},
  \bibinfo{person}{H. Wu}, \bibinfo{person}{S. Zhao}, \bibinfo{person}{D. Yu},
  \bibinfo{person}{H. Tian}, \bibinfo{person}{X. Liu}, {and}
  \bibinfo{person}{R. Yan}.} \bibinfo{year}{2016}\natexlab{}.
\newblock \showarticletitle{Multi-view Response Selection for Human-Computer
  Conversation}. In \bibinfo{booktitle}{\emph{EMNLP}}.
\newblock


\end{thebibliography}

\end{document}